\pdfoutput=1
\documentclass{mn2e}
\usepackage{mnras_cite}
\usepackage{epsfig}
\usepackage{color}
\usepackage{multirow}

\usepackage{graphicx}
\usepackage{subfigure}

\newcommand{\kband}{{$K_S$-band}}
\newcommand{\dif}{{\rm d}}
\newcommand{\eq}{\begin{equation}}
\newcommand{\qe}{\end{equation}}
\newcommand{\ar}{\begin{eqnarray}}
\newcommand{\ra}{\end{eqnarray}}
\newcommand{\fig}{\begin{figure}}
\newcommand{\gif}{\end{figure}}

\begin{document}

\title[Mock observations with the Millennium simulation]{Mock observations with the Millennium simulation: Cosmological downsizing and intermediate redshift observations}
\author[M.~J.~Stringer, A.~J.~Benson, K.~Bundy, R.~S.~Ellis, E.~L.~Quetin]{M.~J.~Stringer$^{1,3}$, 
A.~J.~Benson$^1$, K.~Bundy$^2$, R.~S.~Ellis$^{1,3}$, E.~L.~Quetin$^1$\\ 
1. Caltech, 1200 E. California Blvd., Pasadena, CA 91125, U.S.A. \\ 
2. Dept. of Astronomy and Astrophysics, University of Toronto, 50 St. George Street, Rm 101, Toronto, ON, M5S 3H4,Canada \\
3. Department of Astrophysics, Keble Rd., Oxford, OX1 3RH, U.K.\\ 
} 
\maketitle

\begin{abstract}

 Only by incorporating various forms of feedback can theories of galaxy formation reproduce the present-day luminosity function of galaxies. It has also been argued that such feedback processes might explain the counter-intuitive behaviour of `downsizing' witnessed since redshifts $z\simeq$1-2. To examine this question, observations spanning $0.4 < z < 1.4$ from the DEEP2/Palomar survey are compared with a  suite of equivalent mock observations derived from the Millennium Simulation, populated with galaxies using the {\sc Galform} code. Although the model successfully reproduces the observed total mass function and the general trend of  `downsizing', it fails to accurately reproduce the colour distribution and type-dependent mass functions at all redshifts probed. This failure is shared by other semi-analytical models which collectively appear to  ``over-quench'' star formation in intermediate-mass systems.  These mock lightcones are also a valuable tool for investigating the reliability of the observational results in terms of cosmic variance.  Using variance estimates derived from the lightcones we confirm the significance of the decline since $z \sim 1$ in the observed number density of massive blue galaxies which, we argue, provides the bulk of the associated growth in the red sequence.  We also assess the limitations arising from cosmic variance in terms of our ability to observe mass-dependent growth since $z \sim 1$. 
 \end{abstract}

\section{Introduction}\label{Introduction}

The physical picture of how galaxies assemble has changed markedly over the past decade. A pure `hierarchical dark matter model' in which gas cooling and subsequent star formation occurs in synchronisation with the growth, via gravitational instability, of  their parent dark matter halos fails to reproduce the local luminosity function of galaxies \cite{Benson03,Somerville99,Kauffmann99} and has been challenged by the presence of massive ($\simeq10^{11}\,M_{\odot}$) galaxies at redshifts $z\simeq$2 \cite{Glazebrook04,Cimatti04,vanDokkum06}. As a result, a new paradigm has emerged which argues for the importance of  `feedback' processes that serve to govern the star formation rate in a galaxy. As the efficacy of these processes depends on the mass of the host galaxy, so it is possible to reconcile the predictions of the standard CDM model with the local galaxy luminosity function \cite{Croton06,Bower06}.

Despite this progress, the physical basis of the feedback processes incorporated into the recent semi-analytic models remains largely untested. The most effective way of suppressing star formation and hence inhibiting further growth in massive galaxies is ``radio mode'' feedback \cite{Croton06,Bower06}, where additional gas cooling in halos in which the cooling time is longer than the dynamical time is prevented by low levels of accretion onto central supermassive black holes. \scite{Bower06} have argued that such a process can lead naturally to a characteristic mass scale associated with the transition between cooling on a hydrostatic timescale and more rapid cooling.  Although such a feedback mode can be arranged to match the break in the present day luminosity function, a key issue is whether it explains the trajectory of star formation in galaxies over the past 5-10 Gyr.

Similar progress has been made observationally in measuring the evolving
stellar mass function of galaxies over 0$ < z < $1.5, where large and
complete samples can be obtained
\cite{Fontana04,Drory04,Bundy06,Borch06,Pozzetti07}.  The advent of
large-format near-infrared detectors used in conjunction with deep,
spectroscopic and multi-wavelength surveys has characterized the
evolving stellar mass function \cite{Fontana04}, illuminated the bimodal
nature of local galaxies \cite{Kauffmann03}, demonstrated the presence
of morphological evolution associated with assembly since $z \sim 1$
\cite{Brinchmann00} and revealed how the quenching of star formation
in massive galaxies produces the downsizing signature \cite{Bundy06}.

The time is therefore ripe for a direct confrontation between 
recent simulations which incorporate ``radio mode'' feedback to fit
the local luminosity function and the history of mass assembly over
$0<z<1.5$ from the new generation of deep surveys. By using stellar mass 
estimates as a bridge between theory and observations we can gain 
significant insight not only into the success of the physical prescriptions 
employed by these models \cite{Kitzbichler06}, but also the utility of 
observations in answering the questions posed above.  

A key issue in comparing theoretical predictions with observations is the 
reliability of the latter. Previous comparisons \cite{KW07,Bower06,Cole01} 
have been content to compare the {\it total} stellar mass function. However, the 
signature of downsizing in star formation is most readily tested by examining the
mass function partitioned into star-forming and quiescent populations 
\cite{Bundy06}. While their fractional contributions are relatively easily 
measured, the absolute numbers of these sub-populations are less 
certain because of the limitations of cosmic variance.  This must be properly understood in any comparison, whether
it be theory versus data or one dataset against another.

As an example of the uncertainties arising from cosmic variance, we note 
that while it is generally accepted in the community that the number of 
red-sequence galaxies grows with time, confusion remains over whether 
this growth arises at the expense of a decline in the blue population
\cite{Bundy06}, or is dominated by ``dry'' mergers occurring within the
red sequence, leaving the population of blue galaxies mostly invariant 
\cite{Faber07}. In observational samples broken by redshift bin and galaxy class, 
the impact of cosmic variance becomes a critical issue, sufficient perhaps 
to explain differences in empirical interpretations. In this paper we
will use numerical simulations not only to test popular models of feedback
against observations, but also to evaluate rigorously the limitations
in the data arising from cosmic variance.

Our work follows logically from the earlier study of \scite{KW07}. Those
authors considered the output of lightcones drawn from semi-analytical
models incorporating radio mode feedback as applied to the Millenium 
Simulation \cite{Springel05a} and  compared predictions with various
observables over $0<z<5$. They concluded broad agreement except in 
the abundance of high-$z$ galaxies with large stellar masses. In this paper, 
we will focus on how star formation is distributed within the evolving
galaxy population.

We construct an ensemble of 20 {\em lightcones} (\S\ref{Lightcones}) drawn 
from the {\sc Galform} semi-analytic model \cite{Bower06} as applied to the 
Millennium Simulation.  These lightcones are designed 
to mimic the Palomar/DEEP2 survey observations presented in \scite{Bundy06} 
which we argue in \S2 below is currently the best dataset for addressing the questions of
differential mass assembly and evolution of star forming and quiescent
galaxies.  These lightcones include the detailed survey geometry, optical and 
$K$-band magnitude limits, and photometric as well as stellar mass 
uncertainties described in \scite{Bundy06}. Critically, by studying the 
variation across our 20 realizations, we are also able to determine reliable 
estimates of the effect of cosmic variance. As we will show, this is key to 
an accurate comparison between models and data.

A plan of the paper follows. In \S2 we justify our choice of the DEEP2/
Palomar spectroscopic stellar mass catalog \cite{Bundy06} as the comparison
dataset. In \S3 we discuss the Millenium Simulation and the associated 
{\sc Galform} galaxy formation model. We review the various feedback mechanisms
and illustrate how they lead, in principle, to the concept of downsizing.
We then describe how we construct suitable lightcones and implement
the effect of mass errors in the context of the observational data.
We compare the predicted colour distribution with the observations and
discuss the uncertainties associated with various ways of selecting active
and quiescent galaxies. In \S4 we compare our mass functions with
the observations of \scite{Bundy06}. We also discuss our findings with regard
to how cosmic variance may limit such conclusions, not only in the context
of the current survey \cite{Bundy06} but also in other extant and projected
surveys.
 
\section{Observational Data}

\begin{table*}\label{ObsDetails}
\begin{center}
\caption{A summary of the observational sample (Bundy et al 2006).}
\begin{tabular}{rlrrrrc}
\hline
\multicolumn{2}{c}{Field}&\multicolumn{4}{c}{Number of Galaxies}&Area/deg$^2$\\
\hline
 &&$0.4<z<0.7$&$0.75<z<1$&$1<z<1.4$&Total&\\ 
\cline{3-6}
EGS&\hspace{4.5mm}1&947&1004&722&2673&0.42\\ 
\multirow{3}{*}{DEEP2} & \multirow{3}{*}{$\left\{\begin{array}{c}2\\3\\4\end{array}\right.$} &-&209&31&240&$0.05$\\ 
& &-&353&266&619&0.13\\ 
& &-&644&409&1053&$0.14$\\ 
 \hline
\multicolumn{2}{c}{Totals}&947&2210&1428& 4585 &0.74\\ 
\hline
\end{tabular}
\end{center}
\label{default}
\end{table*}

In this section we review the key features of the dataset presented in
\scite{Bundy06} which will serve as the ``survey template'' for constructing the 
lightcones described in \S3 and the detailed comparisons discussed in \S4.  

In order to derive reliable stellar mass functions, there are two highly desirable 
observational ingredients - extensive multi-band photometry extending 
to the near-infrared, and spectroscopic redshifts for the majority of the 
sample. Although no existing survey has {\it complete} spectroscopic
coverage down to the faintest luminosities probed, we will focus in 
particular on comparisons of the downsizing signature, so good coverage
at the high mass end is certainly advantageous.
 
Our choice to focus on the DEEP2/Palomar catalog is motivated by 
a number of key advantages of this dataset. Foremost, as it builds
on the extensive well-sampled DEEP2 Galaxy Redshift Survey \cite{Davis03}, 
it satisfies the two major criteria above. Furthermore, with respect to the 
important issue of cosmic variance, the survey covers the largest area 
among published surveys with $K$-band imaging---1.5 deg$^2$ spread 
over 4 independent fields. The COMBO17 survey \cite{Wolf03} covers 
$\sim$0.8 square degrees in 3 fields, but has no near-infrared 
photometry.  Although it has no spectroscopic coverage, it does
benefit from highly-calibrated and well-tested photometric redshift data.
The VVDS \cite{leFevre05} data set presented in \scite{Pozzetti07} samples 
only a single field with an area of $\sim$0.5 square degrees, and only 
$\sim$0.2 square degrees has $K$-band imaging.  

We refer the reader to \scite{Bundy06} for further details and only summarize 
the key aspects of the DEEP2/Palomar data set here.  Palomar \kband\ photometry
was obtained in portions of all four fields targeted by the DEEP2 Galaxy
Redshift Survey.  The Extended Groth Strip (EGS) was the top-priority
field and contains the deepest observations.  In \scite{Bundy06} three
different redshift intervals were constructed, 0.4$< z <$0.7, 0.75$<
z <$1.0, 1.0$< z <$1.4 with $K_S$-band depths corresponding to $K_{AB} =$21.8, 22.0 and 22.2 for the three ascending redshift ranges.  
The lowest redshift sample comes entirely from the EGS, since galaxies with 
$z <$0.7 were excluded through colour selection in the other DEEP2 fields.  
In all cases, DEEP2 redshift targets were limited to $R< 24.1$.  The field sizes 
are listed in Table 1.

Central to our later discussion will be the partitioning of this sample
into star-forming and quiescent galaxies. This is key to understanding
the rate at which feedback suppresses star formation and provides
valuable information in addition to the integrated stellar mass function
\cite{KW07}.  \scite{Bundy06} considered both the rest-frame $U-B$ colour 
and the rest-frame equivalent width of the [O II] emission line \cite{Guzman97} 
as proxies for the star formation rate.

\section{Modelling}

Our aim in this work is to construct multiple sets of model galaxies selected in a 
similar manner, and with stellar masses inferred using similar techniques, as for 
the observed galaxies. This will provide the fairest possible comparison between 
theory and data.

\subsection{Mass assembly}\label{Assembly}

The model galaxy samples are generated from the population of dark mater halos in the 
Millennium Simulation \cite{Springel05a}. This simulation consists of approximately 10 billion dark 
matter particles each of mass $8.6\times10^8h^{-1}M_\odot$ evolving in a cubic volume of side 
$500h^{-1}$Mpc, assuming a $\Lambda$CDM cosmology\footnote{Specifically, a flat universe 
with $\Omega_{\rm b} = 0.045,  \Omega_{\rm M} = 0.25,  H_0 = 73{\rm km~s}^{-1}{\rm Mpc}^{-1}$ 
and $\sigma_8=0.9$.}.

Dark matter halo merger trees are found from this 4-volume using the methods 
described by \scite{Harker06}. The lowest mass halos contained in these trees, 
of which there are about 20 million, consist of 20 particles corresponding to a 
total mass of $5\times10^9h^{-1}M_\odot$. Such halos could contain at most 
$9\times10^8h^{-1}M_\odot$ of baryonic material which is well below the lower 
limit of the stellar mass functions to be considered in this work. Therefore we do 
not expect the resolution of the Millennium Simulation to affect our results.

\begin{figure}
\centering
\includegraphics[trim = 8mm 56mm 10mm 138mm, clip, width=\columnwidth]{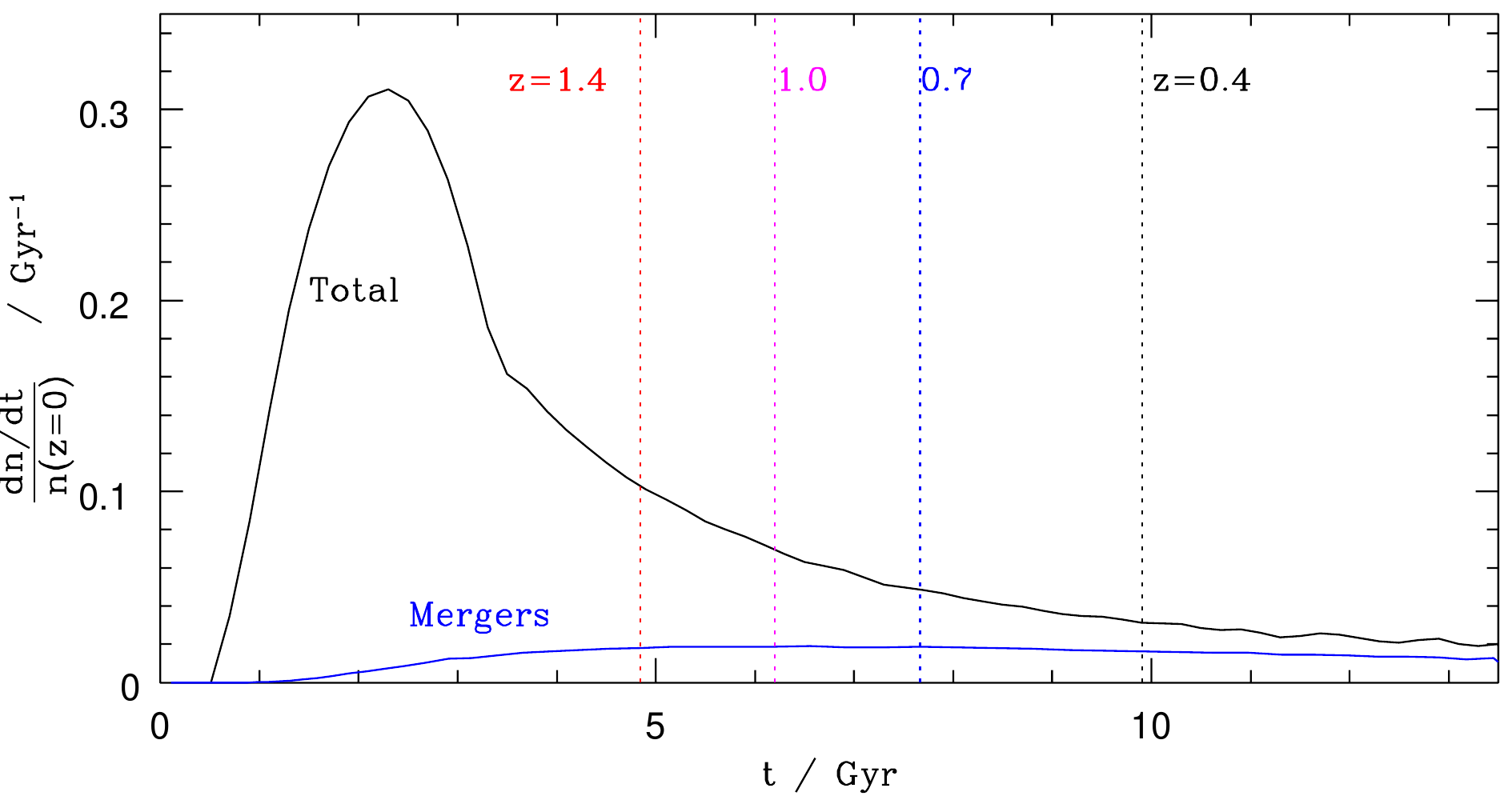}
\caption{The formation rate of galaxies within the Millennium Simulation as a fraction of their present day number. The upper black line shows the total formation rate of galaxies with $M_\star>10^9M_\odot$, the lower blue line shows the rate at which pairs of these galaxies merge with each other. (Mergers involving smaller systems are not included.) The area between the curves will therefore be equal to one. Redshifts relevant to the DEEP2/Palomar samples are indicated with vertical dotted lines.}\label{formation}
\end{figure}

The assembly of dark matter halos in $\Lambda$CDM is often described as ``hierarchical''.  
This is appropriate in that some galaxies from one generation will merge to create the next.  
The importance of this contribution is illustrated in Figure \ref{formation}. The detailed treatment of merging galaxies within a halo can be found in \scite{Bower06}.

The Millenium Simulation demonstrates that the merger rate is expected to be quite small; less than 2\% of galaxies with $M_\star>10^9M_\odot$ merge with each other every Gyr. At early times, minor mergers and the formation of new stars therefore create massive galaxies much faster than they be destroyed by mergers and the latter effect is almost negligible. As the universe evolves, the creation rate of new galaxies diminishes, leading to a near zero net growth in numbers (the difference between the two curves in Figure \ref{formation}). The growth trend is no longer in the number of galaxies but in their individual {\em stellar masses}.
 
\begin{figure}
\includegraphics[trim = 82mm 173mm 8mm 53mm, clip, width=\columnwidth]{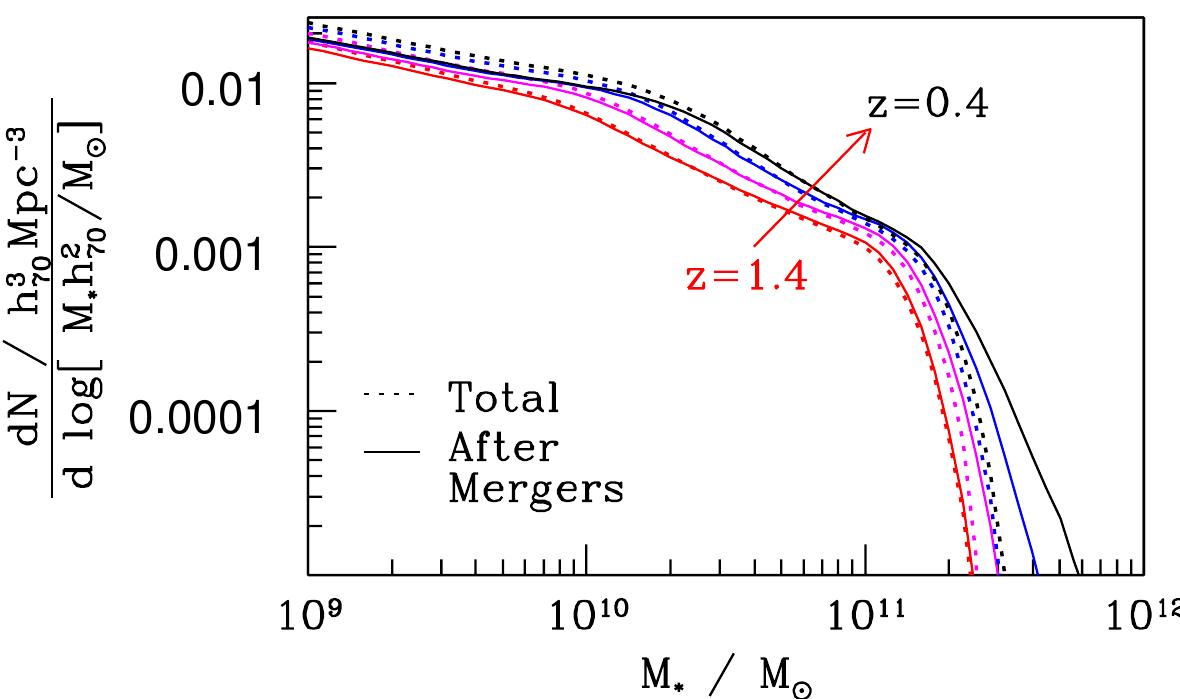}
\caption{The evolving comoving number density of galaxies seen in the Millenium Simulation. Lines are coloured from black to red (top to bottom) corresponding to redshift limits (z=0.4, 0.7, 1.0 and 1.4) adopted in the DEEP2/Palomar survey. The total number of 
systems formed by a given redshift is shown as a dotted line whereas the associated 
solid line shows the true number, including mergers.}
\label{mergers}
\end{figure}

When one considers the {\em differential} mass assembly, therefore, mergers become particularly important. Figure \ref{mergers} shows the total number of galaxies formed up to a given redshift  through accretion (equivalent to the `total' component in Figure \ref{formation}) alongside the number which remain after all mergers (since t=0) have taken place. The latter is the true, final number.

The number density of galaxies in a given mass range will rise or fall depending on 
whether more galaxies arrive or leave that range due to an increase in their stellar 
masses. At intermediate masses, the population  is almost unaffected by mergers; the creation and destruction rates are approximately equal. At high mass, merging has a more significant effect because of the high ratio of potential progenitors to existing galaxies (indicated by a more steeply declining mass function at these masses). A particular deduction is that very massive galaxies ($M_\star>3\times10^{11}M_\odot$) would be almost non-existent without merging. 

The preceding discussion is primarily based upon the formation and interaction of dark 
matter halos in the simulation. However, in order to construct Figure \ref{mergers} we did have to 
consider the stellar content of the galaxies as well as their host halos. We now turn to 
discuss how stellar populations are introduced.

\subsection{Galaxies}\label{Galaxies}

Halos in the merger trees introduced earlier are populated with galaxies using the {\sc Galform} 
semi-analytic model, originally described by \scite{Cole00}. Here we use the 
implementation of that model described in detail by \scite{Bower06}. The reader is 
referred to \scite{Cole00} and \scite{Bower06} for a full description. Here we briefly 
summarise the key physical processes included.

The baryonic component of each dark matter halo is assumed to shock-heat during collapse 
of the halo to the virial temperature, at which point it settles into hydrostatic equilibrium and 
remains pressure supported until it can cool. The radius which has cooled at a given time after 
halo formation is calculated based on the metallicity of the gas, the cooling curve of 
\scite{Sutherland93} and by assuming that it is isothermal at the virial temperature, with 
the radial density distribution:
\eq
\rho_{\rm gas}(r) \propto\left(r^2+r_{\rm core}^2\right)^{-1} ,
\qe 
with $r_{\rm core}$ set equal to one tenth of the virial radius. This choice is motivated by 
the simulations of \scite{Eke98} and \scite{Navarro95}. If the free-fall time for this radius 
has already been passed, any remaining enclosed halo gas is assumed to have settled 
into the central disk, where it becomes rotationally supported by its residual angular 
momentum and the combined gravitational potential of the baryonic and dark matter. 
(The gas is otherwise added to the disk after a free-fall time). 

We note that recent simulations have shown that, in low mass halos, shock heating 
may never occur, with gas arriving into the central galaxy through ``cold flows'' 
\cite{Keres,Birnboim03}. This mode of accretion is correctly accounted for in the model: In such low mass halos the cooling time will become much shorter than the halo dynamical time, and the mass infall rate will become independent of the assumption of shock heating and subsequent cooling and will instead be controlled by the cosmological mass accretion rate of the halo and the  time required for free-fall.

Star formation then proceeds in this cold disk gas with an instantaneous rate, $\psi$ given by:
\eq\label{SFE}
\psi = \epsilon_\star\left(\frac{v_{\rm disk}}{200{\rm kms}^{-1}}\right)^\frac{3}{2}\frac{M_{\rm gas}}{\tau_{\rm disk}},
\qe
where $\tau_{\rm disk}$ refers to the dynamical time in the disk, $v_{\rm disk}$ the circular 
velocity at the half-mass radius and the dimensionless efficiency parameter, $\epsilon_\star=0.0029$.

The key necessity for feedback arises because the above assumptions lead to 
overly rapid cooling and star formation and hence to a galaxy mass distribution 
incompatible with observations \cite{Benson03}. Various physical mechanisms have 
been proposed to provide the necessary feedback effects to prevent this from 
occurring. Two processes are particularly important:

\fig
\includegraphics[trim = 92mm 167mm 9mm 50mm, clip, width=\columnwidth]{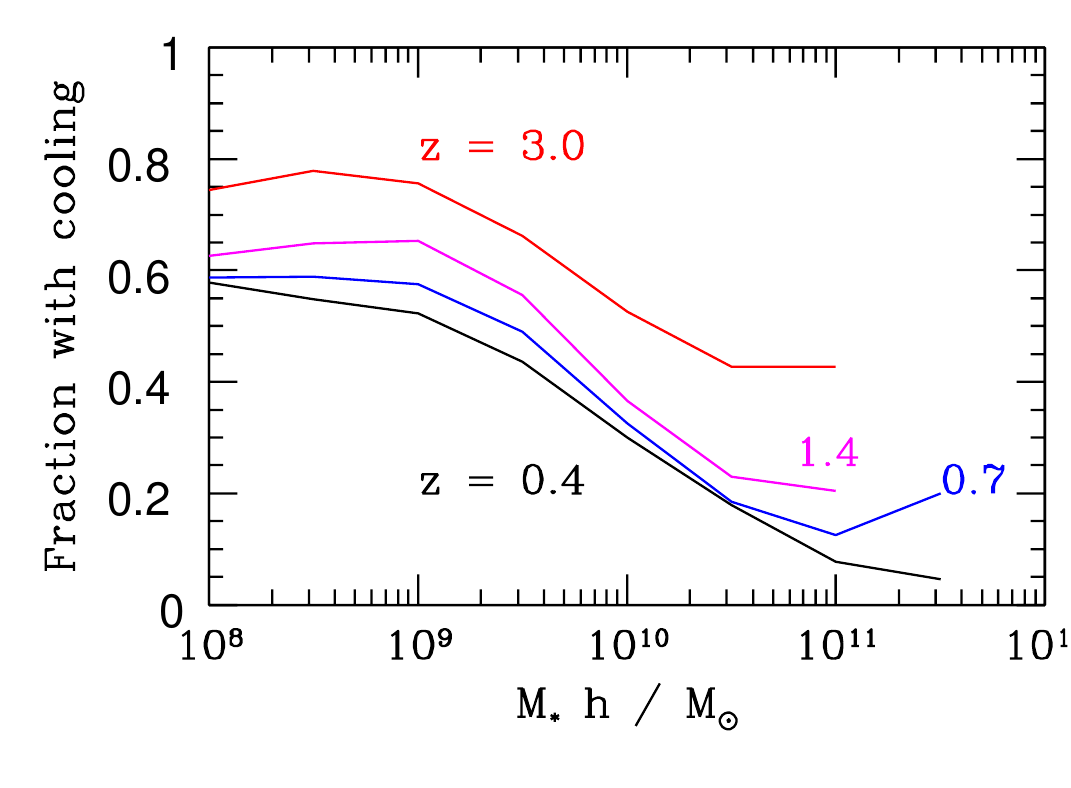}
\caption{The fraction of galaxies with non-zero mass accretion due to
  cooling, plotted as a function of stellar mass, for four redshifts.
  The bottom three span the range of the DEEP2/Palomar
  samples. Significant cooling with subsequent star formation occurs for
  all masses at high redshift, declining to be almost completely
  suppressed for $M_*>10^{11}M_\odot$ by $z\simeq$0.4. This
  demonstrates how the continued suppression of star formation due to
  radio mode AGN feedback can reconcile downsizing in the context of
  hierarchical models.}
\label{CoolingFraction}
\gif

\begin{itemize}
\item{{\bf Reheating}} of disk gas preferentially suppresses the formation of low-mass 
galaxies because of their shallow potential wells. Some fraction of the disk gas is 
reheated by stellar winds and supernovae and is instantaneously (relative to the 
other timescales within the model) returned to the halo. This process is parameterised 
as follows:

\eq\label{Outflow}
\dot{M}_{\rm out} = \beta\psi,\hspace{2cm}\beta= \left(\frac{v_{\rm hot}}{v_{\rm disk}}\right)^{3.2}
\qe
with $v_{\rm hot}=485$km/s. $\beta$ can become extremely large, exceeding $10^4$ 
for small systems and approaching unity only for the most massive galaxies\footnote{These 
values for $\beta$ are extraordinary  when viewed in the context of the physics thought to 
be involved in this outflow process. If eqn.~(\ref{Outflow})  is taken to be energy conserving, 
so that the gravitational potential energy gained by the expelled gas comes directly from 
the stellar winds and supernovae, the energy acquired by the gas per star formed 
would be 
\eq
E = \beta^{3/8}m_\star\left(485{\rm kms}^{-1}\right)^2  .
\qe
For a typical IMF approximately one supernova is expected for every $100M_\odot$ 
of stars formed. This  implies supernovae energies of about
\eq
E \approx\left(\frac{200{\rm km/s}}{v_{\rm disk}}\right)^{1.2}10^{44} {\rm J} .
\qe
This will exceed the available supply for galaxies smaller than the Milky Way, even 
without considering the efficiency with which this is conveyed to the
gas \cite{McKee77}.}.  This  effect cannot, therefore, be ignored when discussing the 
efficiency of star formation. In fact, the process of reheating as described by 
eqn.~(\ref{Outflow}) is ultimately responsible for  determining the eventual 
star formation rate and not, as might be initially expected, the star formation process 
as described by eqn.~(\ref{SFE}). This is discussed further in Appendix~\ref{StarFormation}.

\item{{\bf AGN heating}} was introduced by \scite{Bower06} and occurs if the halo 
free fall time is shorter than the cooling time (formally, $t_{\rm cool}>\alpha_{\rm cool}t_{\rm ff}$, where $\alpha_{\rm cool}=0.58$ is an adjustable parameter) such 
that a hydrostatic halo can exist. If this condition is met, further cooling of gas is prevented 
if the Eddington luminosity of the super massive black hole residing at the centre of the 
galaxy greatly exceeds the cooling luminosity ($L_{\rm Edd}>\epsilon_{\rm SMBH}^{-1}L_{\rm cool}$, where $\epsilon_{\rm SMBH}=0.04$ is an adjustable parameter). This 
strongly suppresses the formation of the most massive galaxies and imprints a near 
exponential cut-off in the abundance of the brightest galaxies. The two free parameters, $\alpha_{\rm cool}$ and $\epsilon_{\rm SMBH}$, were chosen by \scite{Bower06} by constraining the model to match local luminosity function data.

The black hole itself grows through gas accretion triggered by mergers and disk instabilities, 
acquiring $F_{\rm BH}=0.5$\% of the available gas in each merger event and, similarly, 0.5\% of the 
available gas in the galaxy if the disk's self-gravity exceeds the critical limit,
\eq
\frac{{\rm G}M_{\rm disk}}{r_{\rm disk}} > 0.8V_{\rm max},
\qe 
thereby causing the disk to become unstable. This criterion follows the work of 
\scite{Efstathiou82} though the particular constant, and the value of $F_{\rm BH}$, was found by requiring 
the model to match the Magorrian relation between bulge mass and black hole 
mass, $M_{\rm BH} \sim M_{\rm bulge}^{1.12}$, as observed by \scite{Haring04}. 

AGN heating is expected to primarily suppress the formation of the most
massive galaxies.  As a consequence of hierarchical growth, such systems will have
considerable spheroidal components and, hence, massive central black
holes. A study of AGN feedback implemented in semi-analytic models has
been made by \scite{Croton06}, also using the Millennium Simulation, who
found qualitatively similar results.

Figure~\ref{CoolingFraction} illustrates the effect of AGN feedback\footnote{
Figure 7 of  \scite{Croton06} illustrates a related point by showing the effect 
of AGN feedback in their model on the mean behaviour of all galaxies rather 
than the fraction which is affected.} in a simple way by showing the fraction 
of galaxies in the model with active cooling (i.e. those whose cooling has {\em not} 
been shut down by AGN heating). Cooling is largely unaffected in low-mass 
systems but is completely suppressed in the majority of massive systems. This 
illustrates how hierarchical structure formation can be made consistent with the 
observational phenomenon of `downsizing'.

\end{itemize}

In summary, in order for a galaxy to form a significant number of stars, there must 
be a supply of cooling gas from the surrounding halo to counteract the reheating of 
gas by the energy released. In this case equilibrium will quickly be reached which, 
under our particular parameterisation of the processes involved, results in a roughly 
constant specific star formation rate\footnote{Specific star formation rate 
$\equiv \dot{M}_\star / M_\star$.} across all masses (see Fig.~\ref{Efficiencies}). 
However, this balance is only possible in that fraction of galaxies for which the energy 
from the AGN is insufficient to prevent cooling of halo gas. In the case of no cooling, 
the fuel for star formation will quickly be exhausted, causing the galaxy to fade and 
redden. 

Thus, we can expect the galaxy population to be divided into two populations: 
those where cooling is occurring (high star formation rate, blue colours) and those 
with no cooling (little or no star formation, red colours). The relative abundance of 
each of these categories depends on the galaxies' mass. The key question we wish 
to address is whether the growth and abundance of these two populations, which 
depends crucially on the feedback mechanism, matches that observed.

\subsection{Lightcones}\label{Lightcones}
 
\begin{figure*}
\includegraphics[angle=270, trim = 0mm 0mm 15mm 0mm, clip, width=\textwidth]{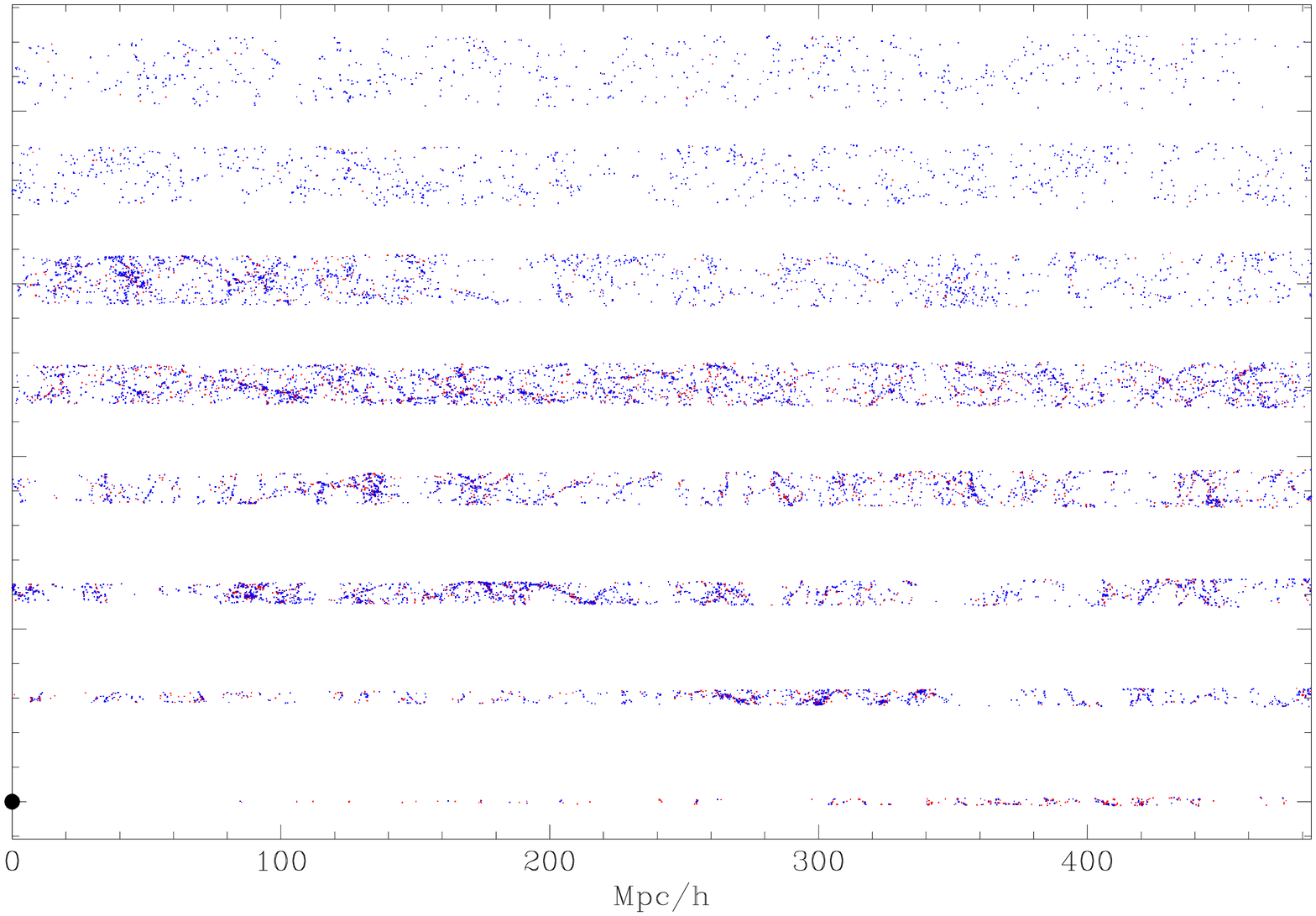}
\caption{Projection of a sample of galaxies on a lightcone drawn from
  the $\sim10^8$Mpc$^3$ comoving volume of the Millennium
  Simulation. To mimic the DEEP2/Palomar survey, galaxies brighter 
  than $m_{\rm R,AB}=24.1$ are selected and split by star formation rate: 
  star-forming ($>0.2M_\odot$/yr; blue) and quiescent ($<0.2M_\odot$/yr; red). 
  The large black circle indicates the location of the observer. The lightcone is
  split into sections to fit on the page. Starting from the observer,
  the lightcones runs from left to right (with distance and redshift
  increasing along the lightcone). At the right hand edge of the page
  the lightcone reappears on the left hand side, shifted up the page by
  60Mpc/$h$. Significant large scale structures---some of which span the
  width of the lightcone---are apparent, highlighting the need for a
  careful analysis of sample variance.}
\label{lightcone}
\end{figure*}
 
We use the prescriptions for gas cooling, star formation, and feedback
described above to populate dark matter halos from the Millennium
Simulation with model galaxies.  The extremely large volume of this
simulation allows us to reduce the statistical uncertainty in our
predictions. However, to assess the significance of potential
differences with observations we must examine various sources of
 uncertainty.  While errors based on photometry or Poisson
statistics are typically straightforward to estimate, the observational
results of interest here---in particular, galaxy number density---are often 
dominated by sample or ``cosmic'' variance, which is much more 
challenging to quantify.  The combination
of a semi-analytic model and the Millennium Simulation provides a
powerful approach to this problem.  After demonstrating general
agreement between simulated and observed results, we construct numerous
mock datasets with the identical field geometry, magnitude limits, and
photometric uncertainties as the real datasets.  By comparing multiple
realizations, we can accurately estimate the effect
of cosmic variance.  Below, we describe our method for constructing such
lightcone datasets\footnote{Such datasets are referred to as
  ``lightcones'' \cite{KW07} as they contain all galaxies which
  intersect the past lightcone of an observer located at some point
  within the simulation volume.}.

We begin by populating the entire Millennium Simulation with galaxies using the 
methods described in \S\ref{Galaxies}. This process is carried out for every 
available redshift in the Millennium Simulation between $z=2$ and $z=0$. In 
addition to physical properties such as stellar mass and star formation rate we 
compute for each galaxy observable quantities such as apparent magnitudes in 
several bands and rest-frame absolute magnitudes.

We then extract lightcones from this multi-redshift dataset using techniques  
similar to those of \scite{KW07}. The only significant difference lies in the method 
of interpolating galaxy properties between the output redshifts of the Millennium 
Simulation as discussed below. Briefly, we select random locations within the simulation 
volume and place an ``observer'' at that point. A line of sight is chosen following the methods of 
\scite{KW07}\footnote{The lightcones constructed have an extent much greater 
than the size of the Millennium Simulation and so use the periodic boundary conditions 
of the simulation to effectively create a larger volume. The methods of \protect\scite{KW07} 
choose the line of sight in such a way as to minimize the possibility of the cone 
intersecting the same region of the simulation in different periodic replications.} 
and a cone with the geometry of the observed sample is constructed around this 
line of sight. We then proceed to identify all galaxies which intersect the past 
lightcone of the observer within this cone.

Kitzbichler \& White compute magnitudes of each galaxy at its output redshift and 
the two adjacent output redshifts, thereby allowing them to interpolate to find the 
magnitude at any intermediate redshift. They effectively apply a differential $k$-correction
to correct for the difference between the output redshift and the observed redshift 
at which the galaxy intersects the past lightcone. In our approach, we track each galaxy 
from one output to the next. Consequently, we can directly interpolate the 
properties of the galaxy to the precise redshift at which it is observed. 
Importantly, this allows us to include both $k$-corrections and evolutionary 
corrections to magnitudes and to interpolate stellar masses which will change from 
one output to the next.

Twenty lightcones are constructed, each having a solid angle of
$0.42{\rm deg}^2$, equivalent to the largest individual area covered by
any of the three DEEP2/Palomar sub-samples.  Galaxies are selected from
within each lightcone if their $R_{\rm AB}$ and $K_{\rm s}$ magnitudes fall within the same limits as the 
observational sample. The variation in number density across these 20 samples can then be calculated.
  
We create additional sets of lightcones by trimming the originals to match the
solid angles of the smaller DEEP2/Palomar fields which are relevant at $0.7<z<1.0$ 
(see Table~\ref{ObsDetails}). For these redshifts, 4845 mock samples are 
generated, each consisting of four lightcones of the appropriate size, and 
the variation measured across these samples. The estimates of sample variance obtained in this way are therefore conservative; slight underestimates because of correlations between the different lightcones.

\subsection{Stellar mass estimates}
\label{MassErrors}

\fig
\centering
\includegraphics[trim = 95mm 140mm 9mm 50mm, clip, width=0.7\columnwidth]{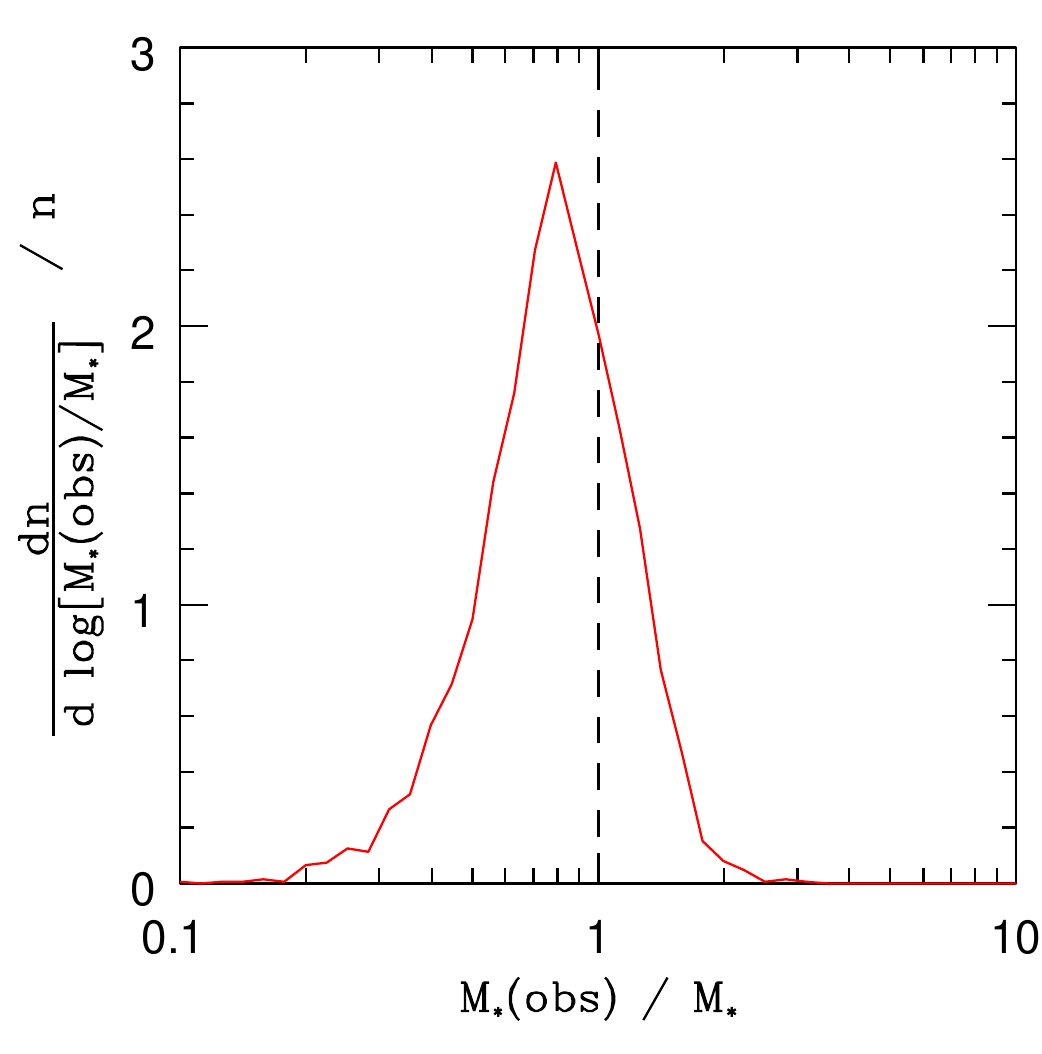}
\caption{The stellar masses of a representative 15,000 galaxies, derived from the 
simulated photometry, plotted as a fraction of the generated stellar mass.  The IMF of \protect\scite{Chabrier03} was used in the model and in the analysis used to make the mass estimates.}
\label{masses}
\gif

Stellar masses in the DEEP2/Palomar survey were derived for each survey
galaxy using the spectroscopic redshift (when available) and a spectral
energy distribution (SED) based on optical and near-infrared
photometry. The effect of possible errors in this process were discussed
in detail by \scite{Bundy06}. In the case of the simulated galaxies, we
can evaluate the uncertainties independently by computing predicted
photometric magnitudes using {\sc Galform} and then applying the same
method used by \scite{Bundy06} on the observed photometry to rederive stellar 
mass estimates.  Comparing these derived masses to the ``input'' values 
determined by the model  provides a useful check on the stellar mass estimates 
and their uncertainties.

For each of 15,000 galaxies, the predicted photometry for model galaxies
was compared to actual data from the EGS field of the DEEP2 survey.
Photometric errors were assigned by randomly sampling the error
distribution of EGS sources with similar magnitudes for each passband.
In this way the perturbed magnitudes of the simulated galaxies reflect
the data quality of the EGS, including the variations in survey depth.
The stellar mass was estimated as in \scite{Bundy06} by comparing the
``observable'' SED of model galaxies to a large grid of stellar
population templates and marginalizing over this grid, the final value
being the median of the mass probability distribution.

The comparison between the derived mass estimates and their input values
are shown in Figure \ref{masses}.  For most galaxies, the agreement is
excellent with a scatter of $\sim$0.15 dex as expected from the analysis
in \scite{Bundy06} of the internal uncertainties of the mass estimator
and photometric errors.  This level of uncertainty is applied to the
model masses when constructing mass functions below.  Our mass
comparison also identified a small subclass of model
galaxies where the rederived mass estimates are larger by
$\sim$0.1 dex than their model values.  These systems are of
intermediate mass and have star formation histories that have been
sharply truncted by the {\sc Galform} model, usually as a result of
entering a larger halo.  The timescales are typically much shorter than
a dynamical time, leading to unphysical stellar populations that are
poorly fit by the stellar mass estimator.  Indeed, these galaxies
populate a region of colour-colour space that is avoided by observed
galaxies and represent the extreme end of the problem of
over-quenching, which we will return to below.  

\section{Results}
\label{Results}

Our primary goal is to compare the downsizing trends claimed by \scite{Bundy06} via
their colour-dependent stellar mass functions and to address the extent to which their
evolving mass threshold is consistent with feedback models which reproduce
the local mass and luminosity function \cite{Bower06,Croton06}. 

\subsection{Colour bimodality and the quenching of star formation}\label{bimodality}
\begin{figure}
\includegraphics[trim = 82mm 55mm 10mm 52mm, clip, width=\columnwidth]{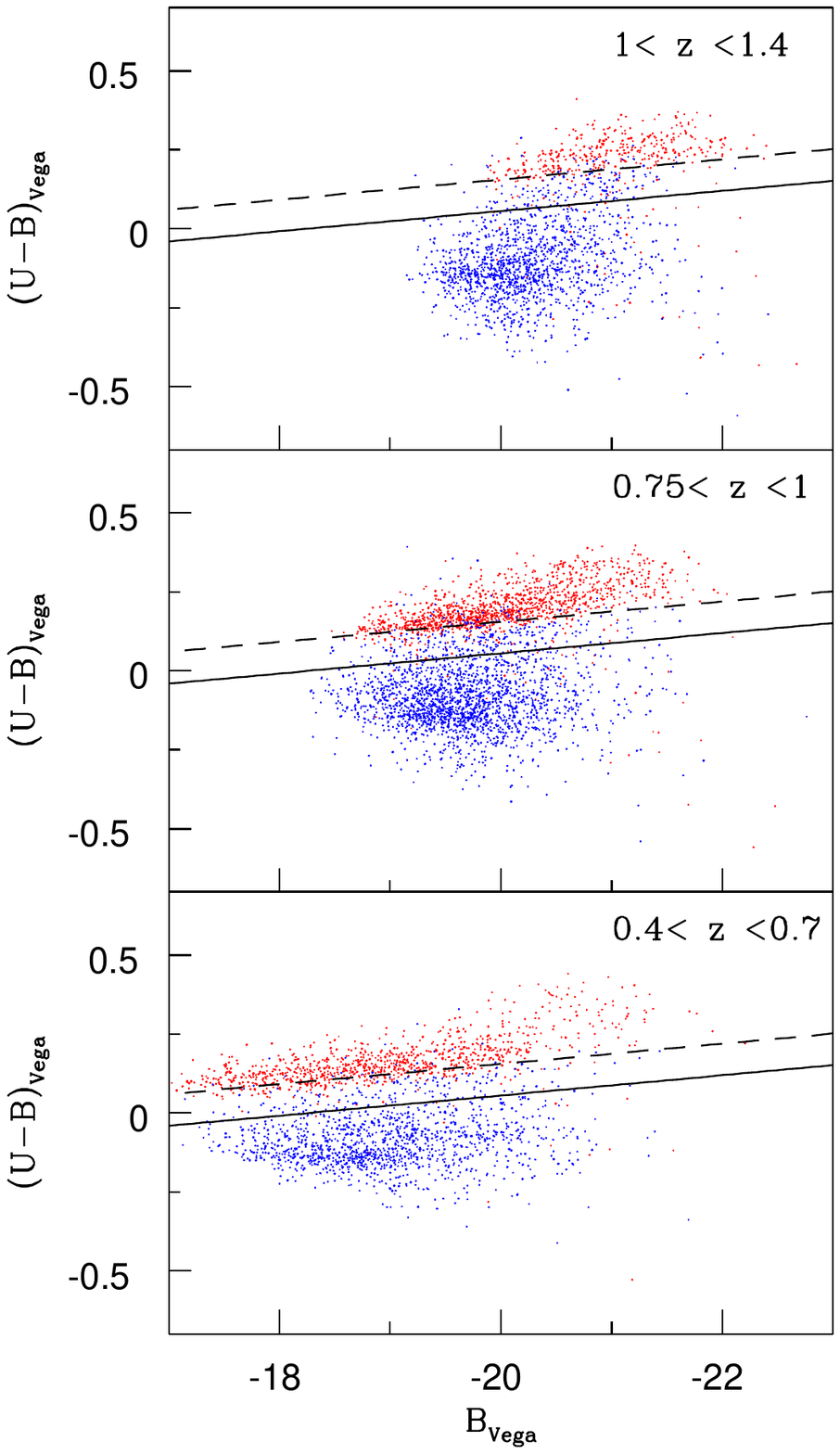}
\caption{Galaxies from one of the 20 lightcone samples. As in \protect\scite{Bundy06}, the 
star-forming and red-sequence galaxies are highlighted by applying a cut on the star 
formation rate at $\dot{M}_\star = 0.2M_\odot{\rm Gyr}^{-1}$. Galaxies above this cut are 
coloured blue while those below are coloured red in Figure \ref{colour_mag}. This criterion 
serves to illustrate the presence of two distinct populations but it does not identify 
each precisely. The solid diagonal line indicates the division made in the {\it observational} sample 
using rest-frame $U-B$  colours. The solid line shows the division in the {\it model} sample, thereby producing Figures \protect\ref{noerrors} \& \protect\ref{complete}.}\label{colour_mag}
\end{figure}
\begin{figure}
\includegraphics[trim = 84mm 55mm 15mm 33mm, clip, width=\columnwidth]{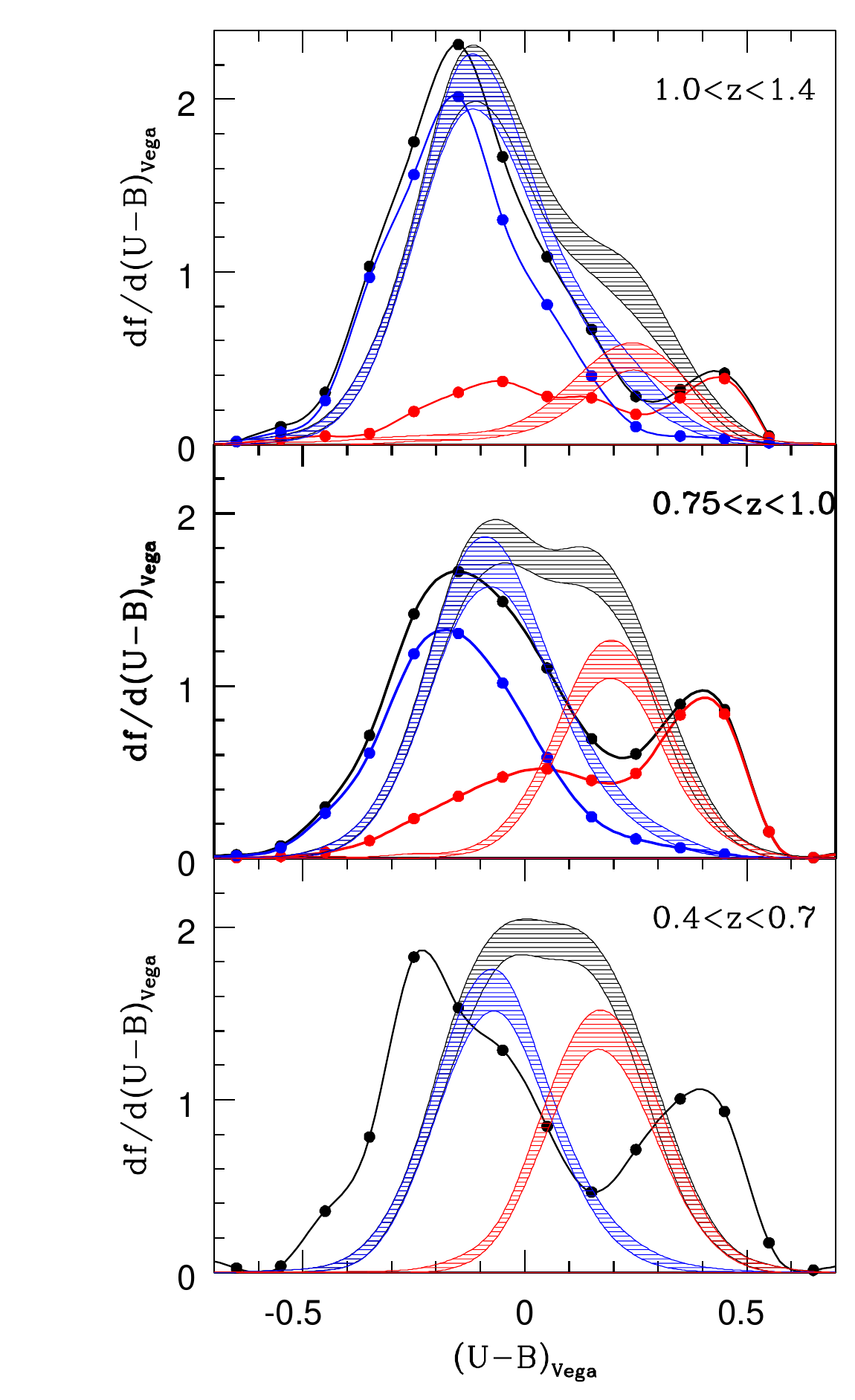}
\caption{The rest-frame $U-B$ colour distribution. Black points show observational 
values  and shaded areas correspond to the total range of values found across 20 
lightcones. For the latter, an observational error of 0.1 mag (1 s.d.) has been included. 
Both datasets are connected with smooth curves for visual clarity. The same star formation rate cut 
as Figure \ref{colour_mag}, $\dot{M}_\star = 0.2M_\odot{\rm Gyr}^{-1}$, is applied to 
illustrate the division into two populations.}
\label{colours}
\end{figure}

We begin our comparison between the DEEP2/Palomar observations and the
{\sc Galform} model with an analysis of the colour distribution.  Figure
\ref{colour_mag} plots in three redshift intervals the restframe ($U-B$) colour-magnitude 
diagram of galaxies drawn from one set of mock lightcones constructed to match the 
properties of the DEEP2/Palomar dataset.  As discussed in \scite{Croton06} and 
\scite{Bower06}, Figure \ref{colour_mag} shows that the incorporation of radio-mode 
AGN feedback helps establish a distinct red sequence and prevents star formation in 
the brightest galaxies, as observed.  The solid line in these diagrams has a slope 
given by \scite{vanDokkum06} with a vertical offset chosen empirically by \scite{Willmer06} 
and \scite{Bundy06} to divide the red sequence from the blue cloud in the 
observed DEEP2 distribution.  This line has the equation:
\begin{equation}
U_r-B_r = -0.032\left(B_r+21.5\right) + 0.454 -0.25
\label{eq:ccut}
\end{equation}

Clearly the colour cut defined by the observational data appears sub-optimal to split 
the model sample. A more appropriate cut (solid line) is therefore used in the investigation into the evolution of these two populations (\S\ref{Evolution} and \ref{colour_mf}). We note that this adjustement did not significantly change the relative fraction of red and blue galaxies, since the bulk of the red population remain above the dotted, observational line in Figure \ref{colour_mag}. 

A more profound discrepancy between the observed and modelled colours is shown in 
Figure \ref{colours} which plots the ($U-B$) distribution in the same three redshift intervals.  
Results from the lightcones are illustrated by shaded curves while data points signify 
the observed distributions from the DEEP2/Palomar dataset.  Beginning with the total 
distribution, we find that model galaxies trace a narrower range in ($U-B$) colour 
than observed galaxies.  Even after including photometric uncertainties of 0.1 mag,  the 
distribution of colours is simply too narrow, particularly at low redshift.

Figure \ref{colours} reveals further insight into the problem that the {\sc Galform} model has in 
reproducing the red sequence.  The total distribution shows evidence for a red sequence 
in the two highest redshift bins, while for $0.4 < z < 0.7$ the colour distribution looks
nearly unimodal.  Even in the two high-$z$ bins the red-sequence is offset blueward of 
the data---as we saw in Figure \ref{colour_mag}---and seems to include larger numbers 
of red systems than observed.  

\pagebreak

\subsection{Evolution of the total mass function}\label{Evolution}

\begin{figure}
\includegraphics[trim = 82mm 55mm 10mm 165mm, clip, width=\columnwidth]{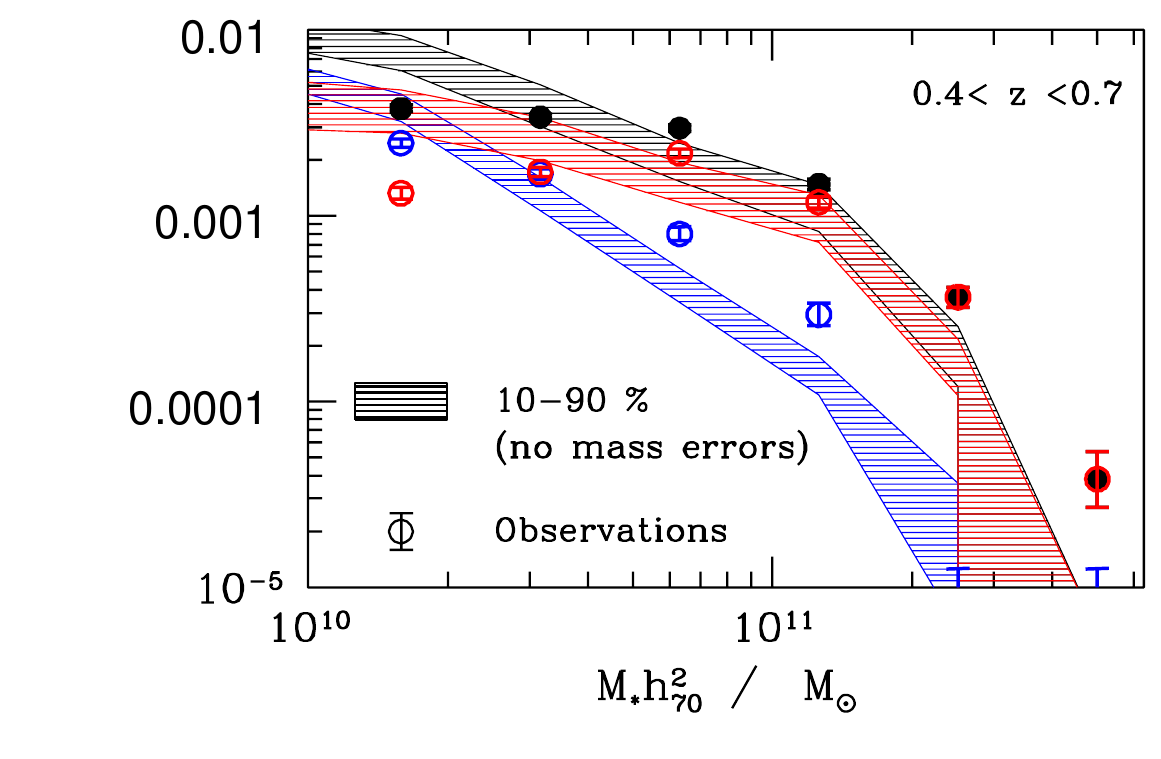}
\caption{The effect of cosmic variance on the derived stellar mass function 
at $0.4<z<0.7$. Points are observational values from \protect\scite{Bundy06}. Shaded areas  
show the 10-90\% range of the simulated mass functions generated from 20 lightcones, 
defined as discussed in section \ref{Lightcones}. Black points and shading relate to the total mass function; red and blue to quiescent and star-forming components, divided as shown in Figure \ref{colour_mag}. Note that, in this figure, stellar masses are taken directly from the model, with no inclusion of errors arising from photometric uncertainties. The effect of including errors can be seen by comparison with the bottom panel of Figure \ref{complete}.}\label{noerrors}
\end{figure}

\begin{figure}
\includegraphics[trim = 82mm 55mm 10mm 52mm, clip, width=\columnwidth]{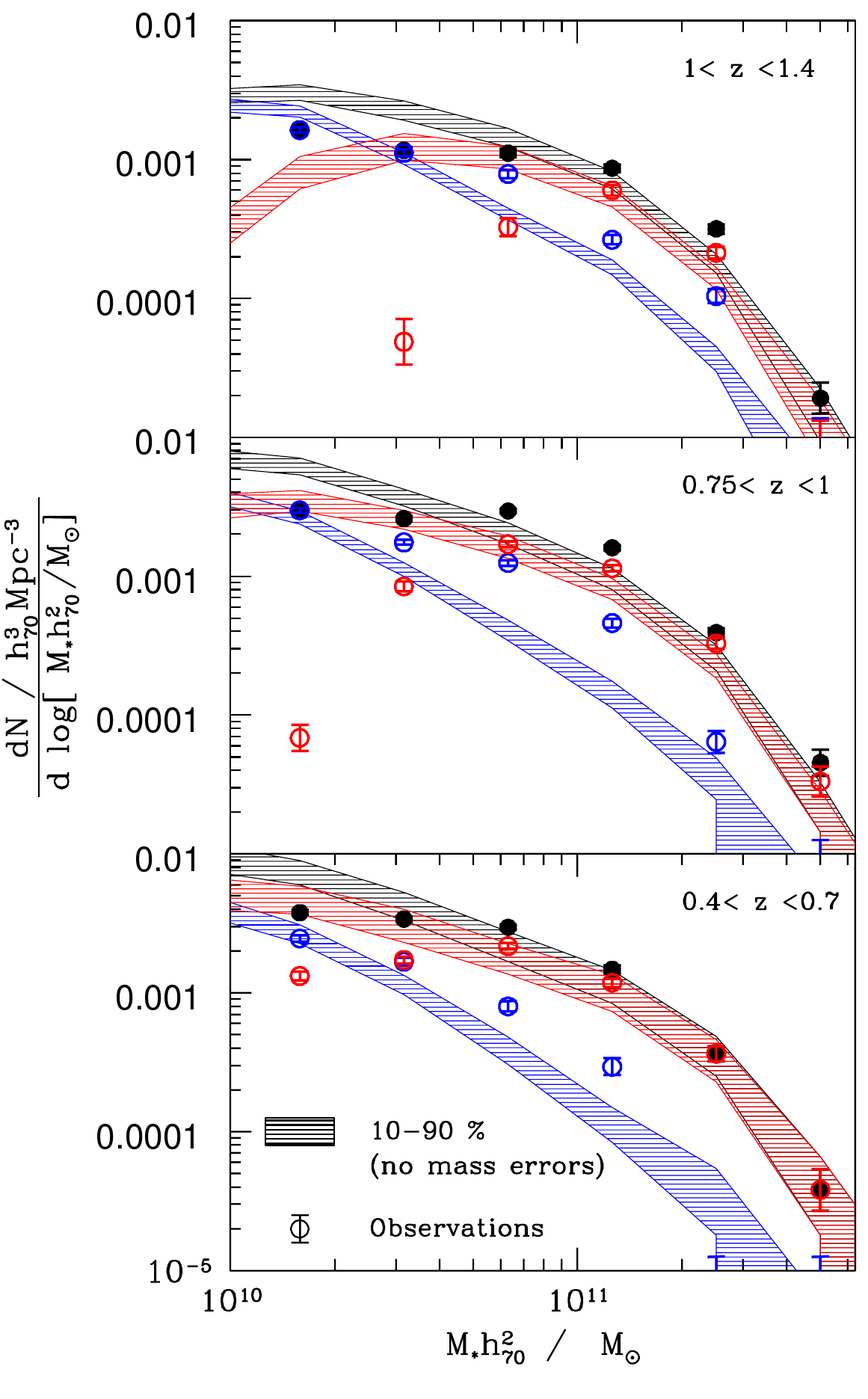}
\caption{The evolving stellar mass function from the DEEP2/Palomar survey compared with 
the output of the semi-analytic models using the {\sc Galform} radio-mode feedback. The 
predictions include the effect of errors in the stellar mass determination. The 10-90\% range 
of the various lightcone realizations are indicated by the shaded areas.  All lightcones are reduced in size so that the four solid angles in each set correspond to those of the four DEEP2/Palomar survey 
fields (see Table \ref{ObsDetails}). The significance of the colours is consistent with Figure \ref{noerrors}.  According to \protect\scite{Bundy06}, the mass completeness limits from the \kband\ alone are $\log [M_\star/M_\odot] = 10.1, 10.2$ and 10.4 for the three respective redshift intervals.}\label{complete}
\end{figure}

Despite the disagreement between the predicted and observed restframe colour 
distributions, we can still evaluate whether the current feedback prescriptions reproduce 
the general mass-dependent decline of star formation in galaxies since $z \sim 1$
which represents the basic `downsizing' signature.  Such a signature should not
be too dependent on the precise definition of what constitutes a quiescent or
active galaxy.

We begin by comparing the total mass function from both the mock lightcones and 
the observations.  The results from the model are plotted in Figure \ref{complete} as 
the black shaded regions while the observations are indicated by black solid points.  The
width of the shading shows the 10-90\% range of values recovered from the lightcone 
samples. As emphasized by \scite{KW07}, we find good agreement with the observed
total mass function {\em when the stellar mass uncertainties are convolved with the model 
results}. This has the effect of increasing the high-mass end of the predicted mass 
function, as can be seen by comparing Figures \ref{noerrors} and \ref{complete}.
We note that we observe fewer low-mass galaxies than predicted at all redshifts, despite the extremely efficient conversion of supernova energy to ejected gas in this model (\S\ref{Galaxies}).   The completeness issues in the observational sample do not explain this discrepancy because the observational magnitude limits have been included in the mock lightcones. 

Though disagreement between the predicted and observed total numbers appears to be minor, this thorough analysis of cosmic variance shows that there is still {\em significant inconsistency} with the data at many stellar mass intervals.

\subsection{Verifying a decline in the star-forming population}\label{colour_mf}

By adopting a suitable division in colour (Fig. \ref{colour_mag}), the model galaxies are divided into red sequence and star-forming populations. It is clear from Figure \ref{complete} that the resulting component mass functions fail significantly to match the observed numbers, even after allowing for cosmic variance and mass errors. However, one important evolutionary trend is qualitatively similar; the red-sequence population growing with time.  The blue population in the model shows no significant evolution however, remaining static across the entire redshift range while the observed numbers fall with time. This leads us to address an important discrepancy of interpretation in the {\em observed} mass functions. 

\scite{Bundy06} argue that massive blue galaxies detected at $z \sim 1$ increasingly transform into quenched red systems, leaving the total mass function relatively
unchanged since $z \sim 1$.  \scite{Faber07} analyze $B$-band luminosity
functions from DEEP2 and COMBO17 and argue that the number of blue
galaxies is essentially unchanging, with the increase in the massive red
sequence population coming primarily from `dry mergers' - mergers of
similarly quiescent galaxies. \scite{Bell07} invoke a third option in
which the red sequence is built primarily from quenched blue galaxies
that are quickly replenished, leaving their numbers unchanged.  The
uncertain effects of cosmic variance have made distinguishing these
scenarios challenging.  With the improved measure of sample variance
afforded by our multiple lightcones, we can now evaluate the validity of
these claims.

The fundamental question is whether the evolution in the mass function
of blue, star-forming galaxies claimed by \scite{Bundy06} is significant
given the uncertainties of cosmic variance.  We have re-analyzed the
observational results of \scite{Bundy06} to derive upper limits on the blue mass
function at the high mass end for the two lower redshift intervals.
These upper limits are determined by the value that would have been obtained had one blue galaxy been detected in each bin.  In all cases, the upper limits are $\dif n / \dif\log(M_*/M_\odot)<10^{-4.9}{\rm Mpc}^{-3}$.

As is apparent in Figure \ref{complete}, this analysis reinforces the
substantial decline observed in blue galaxies---nearly an order of
magnitude across the redshift range at $\log (M_*/M_\odot) \approx 11.4$.
This decrease is significant at the 99.9\% level.  The other
mass bins above the $R$-band determined completeness limit of $\log (M_*/M_\odot)
\approx 10.9$ do not show significant evolution, however.  Using the upper
limits, we find no evolution in the highest mass bin, $\log (M_*/M_\odot) = 11.7$. It should be noted that no blue galaxies
were detected at these masses with $z < 1.0$, so this result is a lower
limit on potential evolution.  At $\log (M_*/M_\odot) = 11.1$, the observations
show an increase from $z \approx 1.2$ to $z \approx 0.9$ followed by an
equally significant decrease from $z \approx 0.9$ to $z \approx 0.5$.
Across the full redshift range, this is consistent with no evolution in
the $\log (M_*/M_\odot) = 11.1$ mass bin at the 60\% level.

Taken together, Figure \ref{complete} and the analysis above substantiates the claim made by \scite{Bundy06} that the population blue, star-forming galaxies observed at $z \sim 1$ declines with time beyond the quenching mass, $\log (M_*/M_\odot) > 11.2$ \cite{Bundy06} . We note, however, that this evolution is not easy to detect even in the large DEEP2/Palomar dataset. Although we see strong and significant evolution in one mass bin, averaging over the relevant mass range reduces the significance to the 2--3$\sigma$ level after cosmic variance.  

Still, such evolution suggests that the red sequence is primarily built from the quenching of star-forming galaxies and only to a lesser extent from dry merging.  A replenishing supply of blue systems would not seem to be necessary since such galaxies are not detected at lower redshifts.  Finally, the substantial uncertainties from cosmic variance, which dominate the error budget, mean that much larger data sets will be required for more detailed studies capable of quantifying the importance of such effects as merging, internal growth due to SF, and transformations on the mass functions of different types of galaxies.  Such work will represent an important step forward in understanding the physical nature of the quenching mechanism.  In Section \ref{variance}, these limitations are discussed in the context of upcoming surveys.

\subsection{Alternative models}
\begin{figure*}
\includegraphics[trim = 8mm 55mm 10mm 52mm, clip, width=1.67\columnwidth]{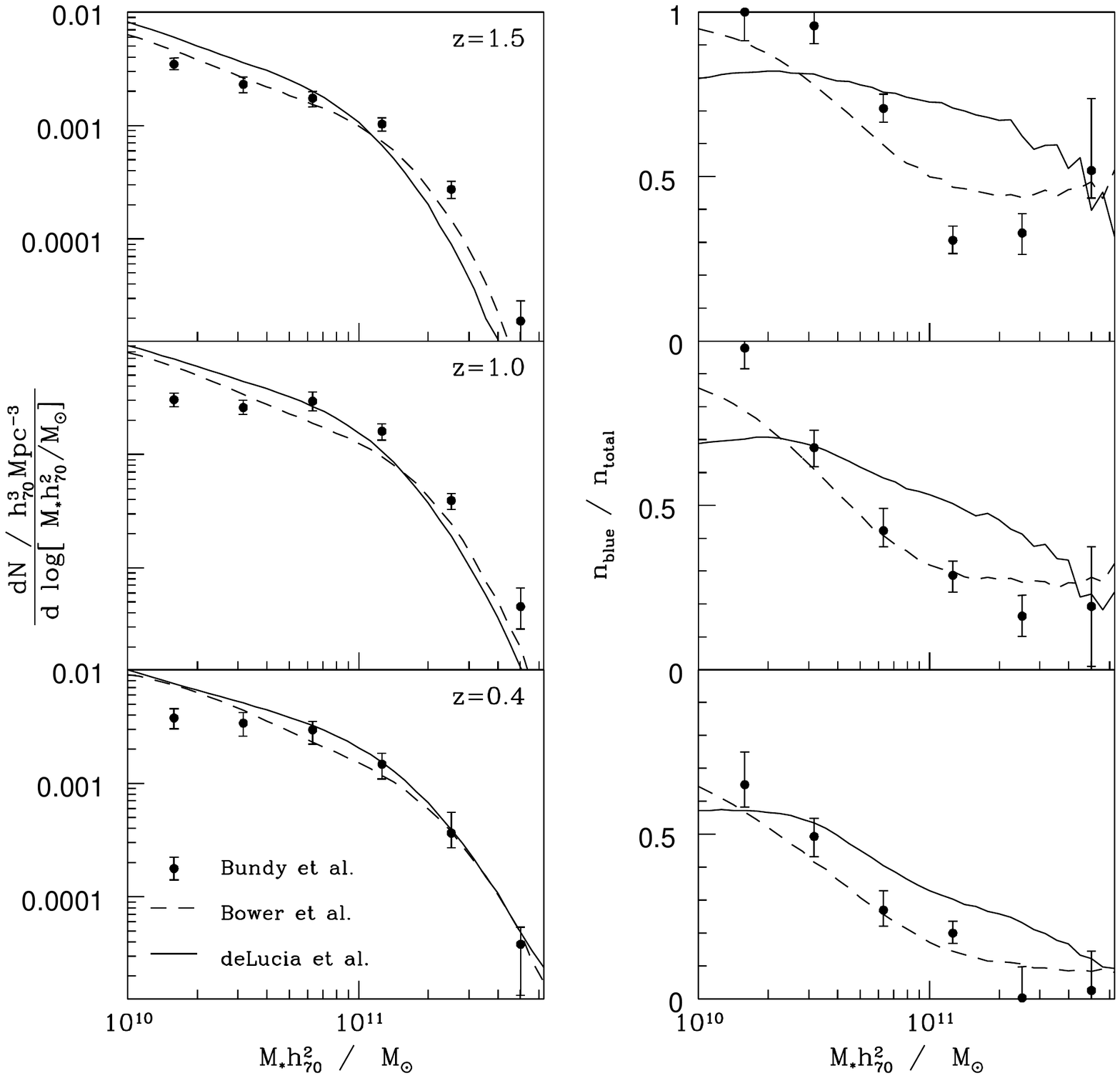}
\caption{Comparison between the stellar mass function predicted by the modelling techique 
and parameters of \protect\scite{Bower06} (dotted lines) and those of \protect\scite{DeLucia07}  (solid lines). Both build on the halo population of the Millennium simulation. They are divided into two populations using the dividing criteria illustrated in Figures~\ref{colour_mag} and \ref{colourVB}. Points show the observational determinations of \protect\scite{Bundy06}, but now with error-bars that reflect the cosmic variance calculated using the techniques of section \protect\ref{Lightcones}. The points are from samples limited in magnitude at R=24.1 (except the highest redshift, which shows the photo-z supplemented mass function from \protect\scite{Bundy06}, limited at R=25.1). Since the model mass functions are derived from instantaneous snapshots from the simulation (not from lightcones) the data is therefore included for reference, not for strict comparison.} \label{deLucia}
\end{figure*}
\fig
\includegraphics[trim = 82mm 55mm 10mm 52mm, clip, width=\columnwidth]{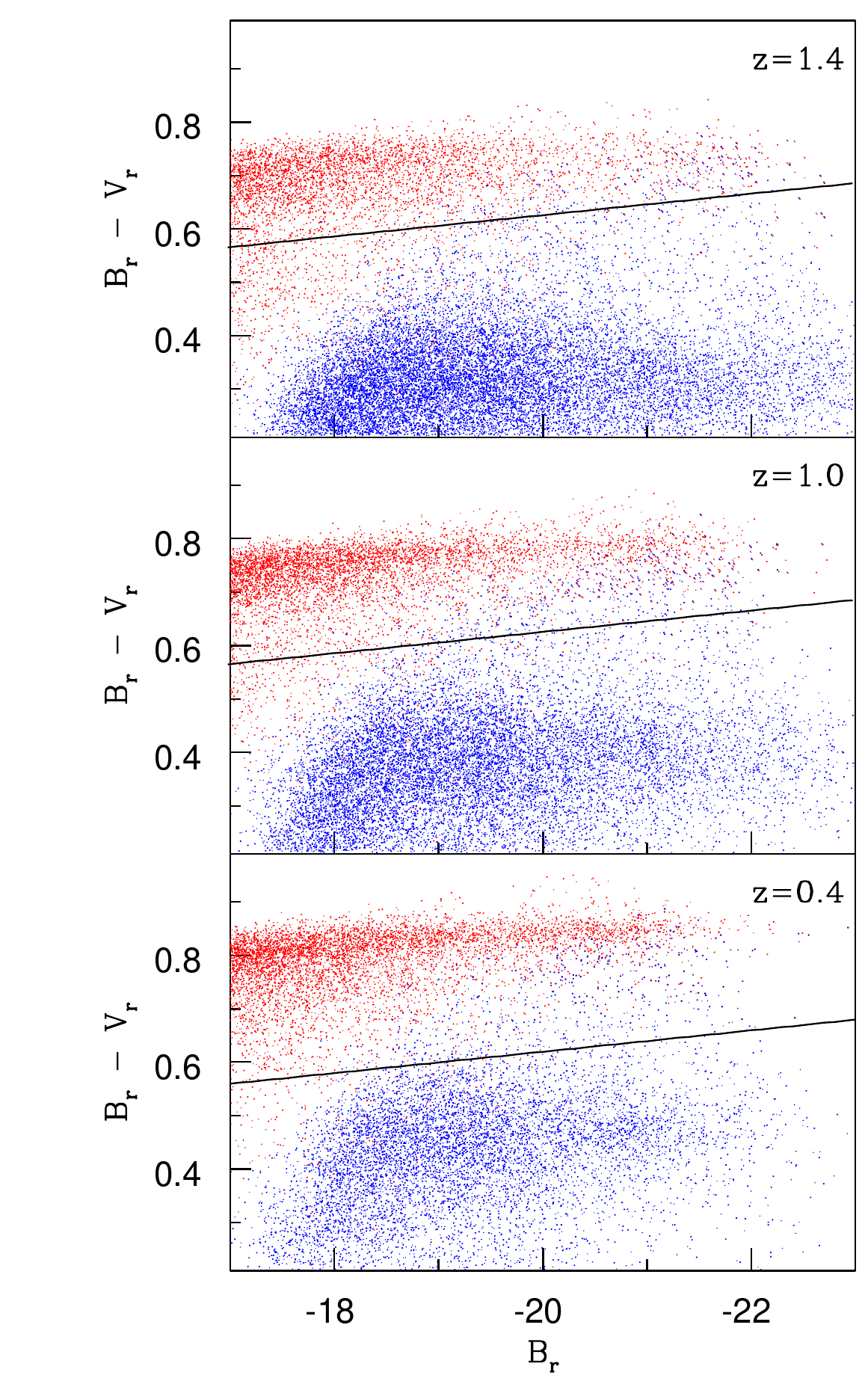}
\caption{The colour magnitude distribution of all the galaxies in a subvolume of the simulation, 
generated using the model of \protect\scite{DeLucia07}. The solid black line indicates the 
colour divide applied to create Fig.~\ref{deLucia}.}\label{colourVB}
\gif
The method applied by \scite{Bower06},  populating the halos of the Millennium simulation using a galaxy formation model, was also adopted by \scite{Croton06}. The results from an updated version of this model \cite{DeLucia07} have been made publically avaiable\footnote{http://www.g-vo.org/Millennium} and the stellar masses and colours of all the galaxies in the simulation volume were extracted.

There are some significant differences in the way that some of the physical processes are modelled. Star formation is assumed to occur only if there is sufficient gas to exceed a critical surface density, following the work of \scite{Martin01}. The mass converted into stars is then 5-15\% of the gas which is above this value. This approach can be compared with equation (\ref{SFE}). Gas is reheated from the disk in line with equation (\ref{Outflow}) but with $\beta=3.5$; the halo potential is involved only when calculating the gas ejected from the halo, not from the disk.

AGN feedback is also incorporated, but rather than imposing a cut-off black hole mass which prevents cooling completely, the cooling rate is simply modified:
\eq
M'_{\rm cool} = M_{\rm cool} - \frac{\eta \dot{M}_{\rm BH}c^2}{\frac{1}{2}v_{\rm vir}^2} .
\qe
The efficiency, $\eta=0.1$. Black hole accretion in this model is also set to match the $M_{\rm BH} - M_{\rm bulge}$ relation mentioned in section \ref{Galaxies}.

The stellar mass function derived from both models is shown in the left hand panel of Figure \ref{deLucia}, for the same redshift ranges as before. It is worth emphasising that these number densities are not derived from lightcone samples, but from the entire volume.  The points from the observational sample are therefore included in Figure \ref{deLucia} for {\em qualitative comparison only}. Detailed quantitative analysis was possible using Figure \ref{complete}, which sets the data against correctly constructed mock observations. 

Despite some considerable differences in their physical assumptions and choice of parameters, it is clear that both models produce very similar predictions. This similarity reflects the fact that those parameters which remain theoretically uncertain are chosen to match each of the models to certain observational data. Similar data sets will have been used in both cases, and will have included the stellar mass function at $z=0$. This partly accounts for the fact that the two models agree more closely with each other (and with the observations) at lower redshift.

In both cases, the predicted stellar mass function evolves in qualitative agreement with these observations. Major disagreement is restricted to distant, low mass galaxies where survey completeness may provide an immediate explaination. 

The colour distribution of galaxies in this second model also shows a clear bimodality (Fig. \ref{colourVB}), as discussed in \scite{Croton06}. The distribution is reproduced here to show the division between red and blue populations which has been used to produce the right hand panel of Figure \ref{deLucia}, showing their fractional abundance.  As with the total number density, both models correctly predict the trend for the blue fraction to decreases with mass and increase with redshift, but specific values are frequently inconsistent with observations. 

The data points in Figure \ref{deLucia} have updated error-bars which now reflect the cosmic variance estimates shown in Figure \ref{complete}. The {\em difference} between the two models is rarely more than this cosmic variance, illustrating an inherent difficulty in constraining competing theories of galaxy evolution on the basis of their predicted stellar mass functions.

\subsection{The effect of cosmic variance on present \& future surveys}\label{variance}

A salutory lesson from Figure \ref{complete} is that even with the largest
and most complete survey of the evolving stellar mass function to date,
cosmic variance is still limiting quantitative conclusions concerning the growth
rate of massive galaxies ($M_\star>10^{11}\,M_{\odot}$) at the level of factors of order 2-3.
These uncertainties have a larger impact than errors in the stellar mass (\S3.4).

Given we have developed the machinery to address this uncertainty, it is interesting
to know the extent to which present and upcoming surveys will be limited
by cosmic variance in addressing basic questions: what is the assembly
rate of the most massive galaxies from $z<2$ to the present day?

\begin{figure}
\includegraphics[trim = 95mm 55mm 10mm 165mm, clip, width=\columnwidth]{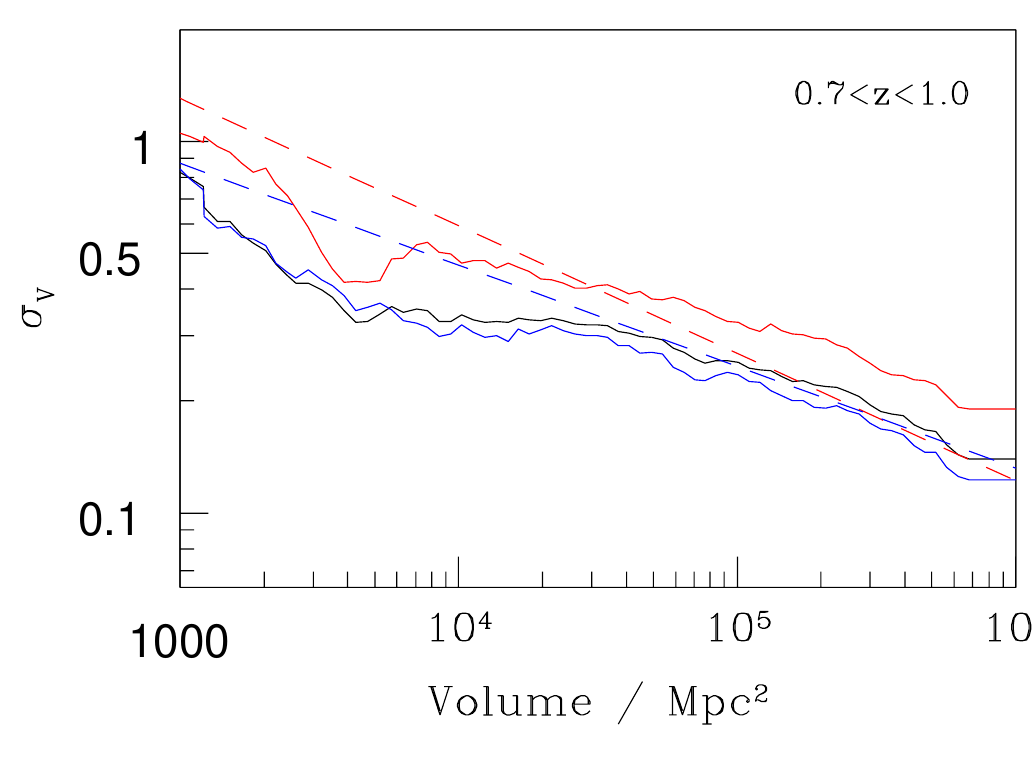}
\caption{The relative cosmic variance of the red and blue galaxy populations as 
a function of survey volume. The solid lines are calculated using the 20 lightcones, the 
dotted lines using the analytical approach of \protect\scite{Somerville04} (see text). The relative variance of the total population of galaxies across the 20 lightcones is shown in black. }\label{somerville}
\end{figure}

In an important paper, \scite{Somerville04} introduced a simple analytical approach
for estimating the effects of cosmic variance in a survey. These analytical estimates  
can be calculated for our mock samples by following the definition used in 
\scite{Somerville04}:
\eq
\sigma_V^2\equiv\frac{\langle N^2\rangle-\langle N\rangle^2}{\langle N\rangle^2} - \frac{1}{\langle N\rangle}.
\qe
It can also be calculated using the correlation length, which is available for the DEEP2 survey \cite{Coil08}. Figure \ref{somerville} compares this analytical prediction with the variance derived
directly from our multiple lightcones technique. As can be seen, the relative agreement
is quite encouraging.

\begin{figure*}
\includegraphics[trim = 7mm 60mm 10mm 45mm, clip, width=\textwidth]{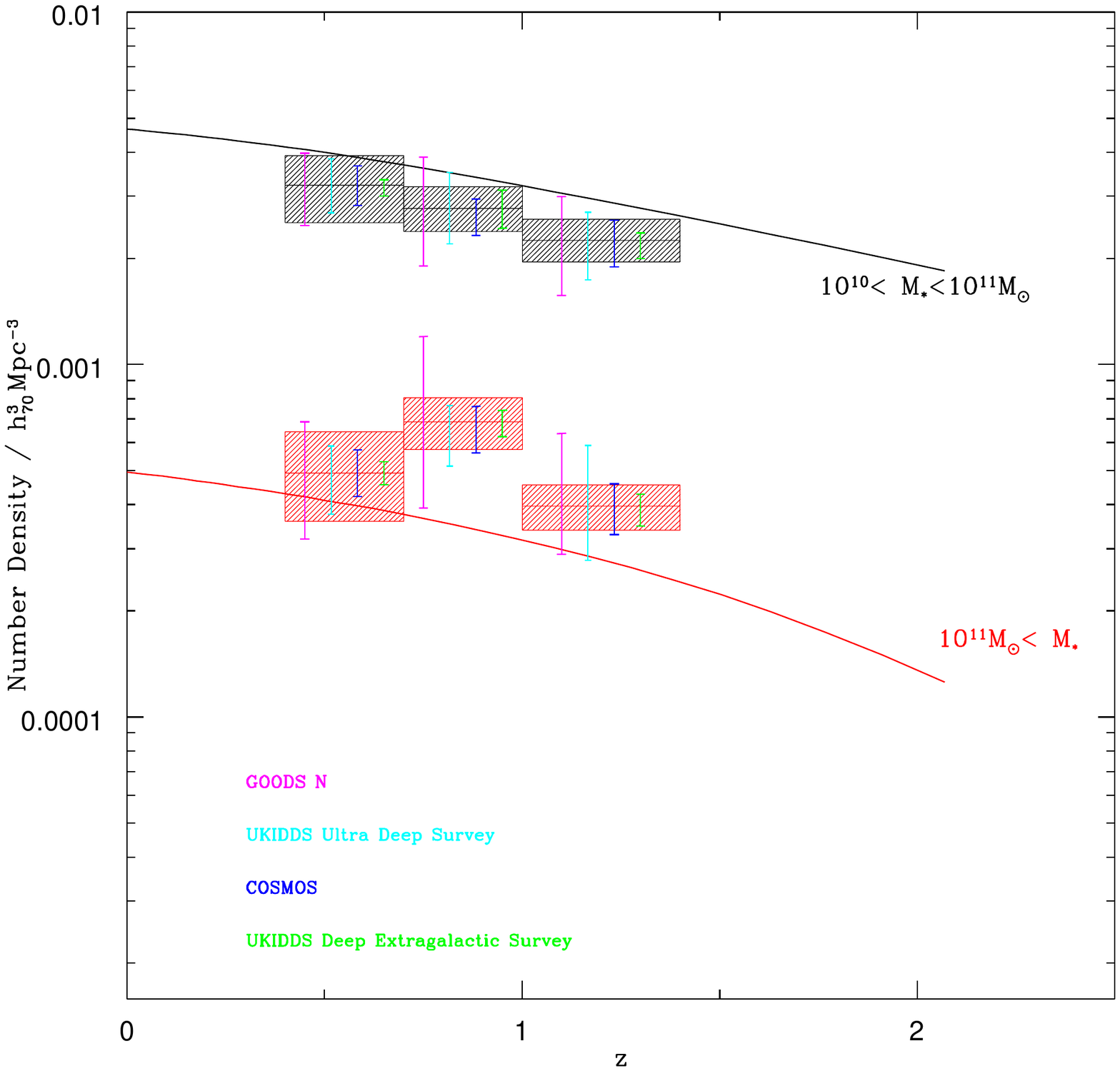}
\caption{The predicted comoving number density of galaxies derived from the full volume 
of the Millenium Simulation for intermediate-mass galaxies (black curves) and those
with high stellar mass (red curves). Shaded areas indicate the constraints
on mass assembly in these two mass bins according to the results of \protect\scite{Bundy06}.
The mean number densities for 10$<\log(M_\star/M_\odot)<$11 may be affected by some incompleteness, which becomes important at $\log(M_\star/M_\odot)<10.4$ in the photo-z supplemented sample used here ($R_{AB}<25.1$).  However, the mock observations confirm that  most of the discrepancy between the data and the models at lower masses is real and not simply the result of incompleteness (Fig.\ref{complete}). The shading indicates 10-90 \% error due to cosmic variance, calculated using the techniques described in \S\ref{Lightcones}. Error bars illustrate the equivalent error expected in other present and upcoming surveys  (left to right are listed from top to bottom). For convenient comparison, these errors are all given relative to the same, observationally determined mean densities. }
\label{assembly}
\end{figure*}

Returning to the physical questions posed above. Figure \ref{assembly} shows the number 
density of intermediate-mass  ($10^{10} < M / M_{\odot} < 10^{11}$) and high-mass 
($M > 10^{11} M_{\odot}$) galaxies for the entire volume of the Millennium Simulation. 
The key question is the extent to which present and upcoming surveys can verify not only the basic trends of mass assembly but also the mass-dependent variations.

First, we consider the situation for the data evaluated in the present paper. 
For the \scite{Bundy06} data plotted in Figure \ref{assembly}, errors correspond 
to the 10-90\% range of values found from the 20 lightcone realizations of this survey. 
Over 0$<z<$1.4, the predicted growth trends in intermediate mass galaxies (upper, black line) are only marginally confirmed and the predicted growth in massive galaxies (lower, red line) is inconsistent with observations. 

Much of the differential trends observed are affected by the area of the survey and the optical magnitude limit (Table 2). The situation with earlier HST samples (GOODS, \scite{Giavalisco04}) is considerably worse. However, even with upcoming panoramic surveys underway with  HST (COSMOS, \scite{Scoville07} ) and UKIRT (UKIDSS, \scite{Lawrence07}), the situation is not significantly better. Thus precise observational constraints concerning the {\em differential evolution} of mass assembly must await a future generation of surveys. 

\begin{table}\label{Surveys}
\begin{center}
\caption{Some details of the selection of observational surveys referenced in Figure \ref{assembly}.}
\begin{tabular}{llrrrc}
\hline
\multicolumn{2}{c}{Field}&Area/deg$^2$&\multicolumn{2}{c}{Mag. Limits}\\
\hline
& & &$R_{\rm AB}$& $K_{\rm vega}$\\ 
\cline{4-5}
\\
\multicolumn{2}{l}{EGS \& DEEP2}&Table 1&24.1&20.5\\
\\
GOODS North& &$1\times 0.042$&24.1&21\\
\\
COSMOS&&$1\times 1.5$&25&22\\
\\
\multirow{2}{*}{UKIDDS} & \multirow{2}{*}{$\left\{\begin{array}{l}{\rm D.E.S.} \\{\rm U.D.S.}\end{array}\right.$}&$4\times 8.75$&23&21\\
&&$1\times 0.77$&27&23\\
\\
\hline
\end{tabular}
\end{center}
\label{default}
\end{table}

\pagebreak
\hspace{4cm}

\pagebreak
\hspace{4cm}

\pagebreak

\section{Summary}

Lightcones derived from the Millenium Simulation, designed specifically to match the geometry and selection parameters of the DEEP2/Palomar survey of Bundy et al (2006), reproduce reasonably well the evolving stellar mass function over 0.4$<z<$1.2. These mock observations are populated with galaxies using the {\sc Galform} code, which incorporates a prescription for `radio mode'  feedback. Such feedback processes are thus an adequate explanation of the broad trend of `downsizing' seen in recent redshift surveys.

However, the {\sc Galform} code can not satisfactorily reproduce the evolving form of the bimodal colour
distribution, or the stellar mass functions of the corresponding two populations. An alternative model \cite{DeLucia07},  proposed originally by \scite{Croton06}, produces similar discrepancies; both models appear to over-quench star formation in intermediate mass galaxies.

This approach also allows us to address the important question of how cosmic
variance may limit the validity of various conclusions drawn from observational surveys.
Using variance estimates derived from the lightcones, we confirm the significance of the
decline since $z\simeq$1 in the number density of massive blue galaxies claimed by Bundy
et al (2006). We argue that the transformation of these blue galaxies must, necessarily, provide
the bulk of the associated growth in red sequence galaxies given the near-constant total
mass function. 

This discussion of cosmic variance is extended to demonstrate the limitations of other,
more ambitious, surveys ongoing at the present time in terms of detecting the {\it mass-dependent}
growth of galaxies since $z\simeq$1.

\section*{Acknowledgments}

We would like to thank Carlton Baugh, Richard Bower, Shaun Cole, Carlos Frenk, John Helly, Cedric Lacey and Rowena Malbon for allowing us to use the {\sc Galform} semi-analytic model of galaxy formation (
{\tt www.galform.org}) in this work. We also thank Simon White for helpful comments.

The Millennium Run simulation used in this paper was carried out by the Virgo Supercomputing Consortium at the Computing Centre of the Max-Planck Society in Garching. The databases and the web application providing online access to them were constructed as part of the activities of the German Astrophysical Virtual Observatory 
{\tt http://www.g-vo.org/Millennium}. 

MJS acknowledges support from the Warden and Fellows of New College, Oxford, the hospitality of the CTCP at Caltech and of the KITP, Santa Barbara. AJB acknowledges support from the Gordon \& Betty Moore Foundation. RSE acknowledges financial support from the Royal Society.

\appendix
\section{Star Formation Efficiency}\label{StarFormation}

In this Appendix we illustrate the importance of gas ejection from the disk, which was given in eqn.~(\ref{Outflow}), for determining the stellar masses of galaxies. The differential equation which controls the mass of cold disk gas, $M_{\rm cold}$, is
\eq\label{m_cold}
\dot{M}_{\rm cold} = \dot{M}_{\rm in}  - (1-R+\beta)\frac{M_{\rm cold}}{\tau_\star}. 
\qe
where $\tau_\star=\tau_{\rm disk}/\epsilon_\star/(v_{\rm disk}/200\hbox{km/s})^{3/2}$. The parameter $R=0.39$ is the fraction of material in stars which will be recycled back into the interstellar medium in the instantaneous recycling approximation\footnote{This particular value is calculated from the initial mass function of \scite{Kennicutt83}:\[\label{IMF}
\zeta(M) \propto  \left\{ \begin{array}{ll} 
M^{-1.4} \hspace{1cm}(0.1&\leq M \leq 1M_\odot) \\
M^{-2.5} \hspace{1cm}(1.0&\leq M \leq 100M_\odot) \end{array}\right. .
\]}. 

For the relevant range of circular velocities, $5<\beta<10,000$, so outflow will be fierce and the only way that a significant supply of cold gas can be retained in the disk is through constant replenishment from the cooling halo gas. This corresponds to the limit $\dot{M}_{\rm cold} \rightarrow 0$ for which equation (\ref{m_cold}) has the following solution:
\eq
M_{\rm cold} \approx \dot{M}_{\rm in}\tau_{\rm eff},\hspace{1.5cm}\left[\tau_{\rm eff} = \frac{\tau_\star}{1-R+\beta}\right] .
\qe
Given the large values of $\beta$, the effective timescale, $\tau_{\rm eff}$ will be extremely short, leading quickly to a steady state where the rates of cooling and reheating gas are approximately constant and equal. In this case, cold gas mass and star formation rate are given by:
\eq\label{Mstar_Min}
M_{\rm cold} \approx \dot{M}_{\rm in}\tau_{\rm eff},\hspace{1.5cm}\dot{M}_\star \approx \dot{M}_{\rm in } \frac{1-R}{1-R+\beta}.
\qe
(Note that the star formation rate given here is the rate of increase of mass in long-lived stars, hence the inclusion of the factor $1-R$.) Because of the tendancy towards these limits, the star formation rate in any particular galaxy is approximately constant (assuming a constant $\dot{M}_{\rm in}$), ignoring short term perturbations resulting from mergers and gravitational instabilities which contribute little over cosmological timescales. In this approximation, the stellar mass is given by the star formation rate integrated over the available time, $t(z)$.

\begin{figure}
\centering
\includegraphics[width=0.8\columnwidth]{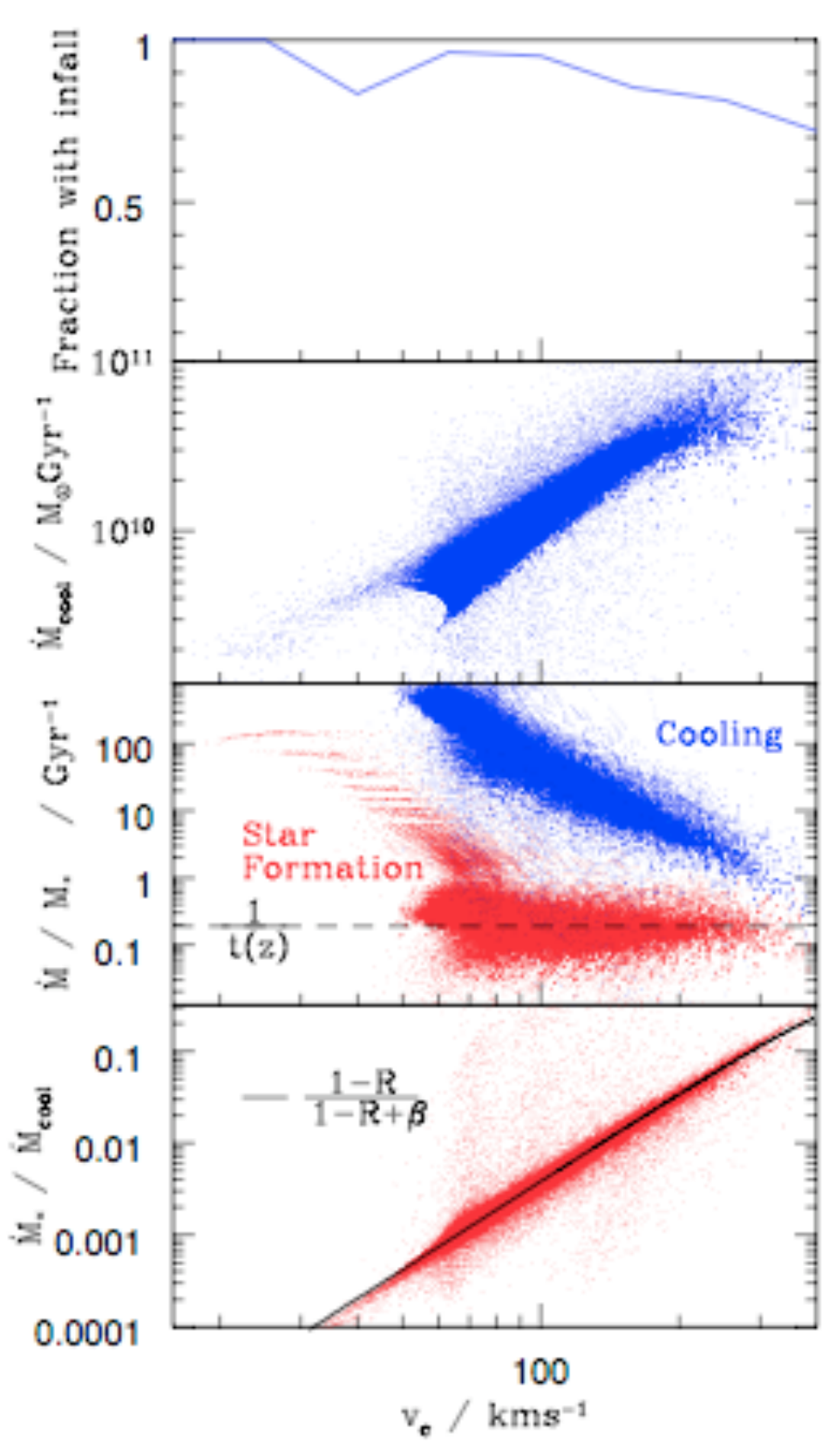}
\caption{An illustration of the relationship between cooling rate and star formation rate as a function of disk circular speed, $v_c$ for $\sim400,000$ galaxies in an arbitrary sub-volume of the model at redshift $z=0.5$. The top panel shows the (weighted) fraction of galaxies which have any cooling at all and so appear on the other three sets of axes. The second panel shows the cooling rate itself and the third panel shows both cooling rate and the star formation rate as a fraction of the existing stellar mass of each galaxy. The dashed line shows the constant star formation rate which would be required to accumulate the relevant stellar mass. The bottom panel illustrates that the ratio between the star formation and cooling is a function of the outflow efficiency, $\beta(v_{\rm c})$, as predicted by eqn.~(\ref{Mstar_Min}). }
\label{Efficiencies}
\end{figure}

The specific star formation rates for galaxies at redshift $z$ should therefore be expected to be scattered about the value
\eq\label{SSFR}
\frac{\dot{M}_\star(z)}{M_\star(z)} \approx \frac{1}{t(z)}.
\qe
Figure~\ref{Efficiencies} demonstrates that these analytic approximations are broadly consistent with the calculated values for $\sim400,000$ galaxies in a sub-volume of the model.

The above argument implies that specific star formation rate is {\em independent} of the equations which explicitly govern it. The parameters for star formation efficiency, $\epsilon_\star$ in eqn.~(\ref{SFE}), and the strength of outflow, $\beta$ in eqn.~(\ref{Outflow}) will have only a secondary influence of the underlying trend described by eqn.~(\ref{SSFR}).

The physical property which {\em is} controlled by the star formation efficiency is the mass of cold gas in the disk relative to the supply from the halo and, consequently, the stellar mass:
\eq\label{EffectiveSFE}
\frac{M_{\rm cold}}{M_\star} \approx \frac{\tau_\star(\epsilon_\star)}{t(z)}.
\qe
So, more efficient star formation leads to a lower gas fraction but {\em not} to higher star formation rates for galaxies of a given mass.

The cooling rate and outflow efficiency, $\beta$, determine the size of the halo in which a galaxy of a certain mass must reside in order to fuel its star formation. This is how the smaller-scale physics of star formation and supernovae influence the collective properties of galaxies on cosmological scales.

\begin{thebibliography}{widestentry}

\bibitem[Baugh et al.<2005>]{Baugh05}
Baugh, C. M., Lacey, C. G., Frenk, C. S., Granato, G. L., Silva, L., Bressan, A., Benson, A. J., \& Cole, S. 2005, MNRAS, 356, 1191

\bibitem[Bell et al.<2007>]{Bell07}
Bell~E.~F., Zheng~X.~Z., Papovich~C., Borch~A., Wolf~C. \& Meisenheimer~K., 2007, ApJ, 663, 834

\bibitem[Benson et al.<2003>]{Benson03}
Benson, A. J., Bower, R. G., Frenk, C. S., Lacey, C. G., Baugh, C. M., Cole, S. 2003, ApJ, 599, 38

\bibitem[Birnboim \& Dekel<2003>]{Birnboim03}
Birnboim, Y., Dekel, A., 2003, MNRAS, 345, 349

\bibitem[Borch et al.<2006>]{Borch06} 
Borch, A., et al.  2006, A\&A, 453, 869 

\bibitem[Bower et al.<2006>]{Bower06}
Bower,~R. G., Benson,~A.~J., Malbon,~R., Helly,~ J.~C., Frenk,~C.~S., Baugh,~C.~M., Cole,~S. \& Lacey,~C.~G. 2006, MNRAS, 370, 645

\bibitem[Bundy et al.<2005>]{Bundy05} 
Bundy, K., Ellis, R.~S., \& Conselice, C.~J. 2005, ApJ, 625, 621 

\bibitem[Bundy et al.<2006>]{Bundy06} 
Bundy,~K., et al. 2006, ApJ, 651, 120 

\bibitem[Brinchmann \& Ellis<2000>]{Brinchmann00}
Brinchmann~J. \& Ellis~R.S., 2000, ApJ, 536, 77 

\bibitem[Cattaneo et al.<2008>]{Cattaneo08}
Cattaneo,~A., Dekel,~A., Faber,~S.~M., Guideroni,~B. 2008, arxiv:0801.1673

\bibitem[Chabrier<2003>]{Chabrier03}
Chabrier, G. 2003, PASP, 115, 763

\bibitem[Cimatti et al.<2004>]{Cimatti04} 
Cimatti,~A., et al. 2004, Nature, 430, 184 

\bibitem[Coil et al.<2008>]{Coil08}
Coil,~A.~L. et al. 2008, ApJ, 672, 153

\bibitem[Cole et al.<2000>]{Cole00}
Cole~S., Lacey~C., Baugh~C. \&  Frenk~C. 2000, MNRAS, 319, 168 

\bibitem[Cole et al.<2001>]{Cole01}
Cole~S. et al. 2001, MNRAS, 326, 255 

\bibitem[Croton et al.<2006>]{Croton06}
Croton, D.~J., et al. 2006, MNRAS, 365, 11 

\bibitem[Davis et al.<2003>]{Davis03}
Davis, M. et al. 2003, Proc. SPIE, 4834, 161

\bibitem[De Lucia \& Blaizot<2007>]{DeLucia07} 
De Lucia, G., \& Blaizot, J. 2007, MNRAS, 375, 2 

\bibitem[Drory, Bender \& Hopp<2004>]{Drory04}
Drory~N., Bender~R. \& Hopp~U., 2004, ApJ, 616, 103

\bibitem[Efstathiou, Lake \& Negroponte<1982>]{Efstathiou82}
Efstathiou G., Lake G., Negroponte J. 1982, MNRAS, 199, 1069

\bibitem[Eke et al.<1998>]{Eke98}
Eke,~V.~R., Cole,~S., Frenk,~C.~S. \& Henry,~J.~P. 1998, MNRAS, 298, 1145

\bibitem[Faber et al.<2007>]{Faber07}
Faber~S.~M. et al. 2007, ApJ, 665, 265 

\bibitem[Fontana et al.<2004>]{Fontana04}
Fontana~A. et al., 2004, AA, 424,23

\bibitem[Giavalisco et al.<2004>]{Giavalisco04}
Giavalisco, M. et al., 2004, ApJ, 600, L93

\bibitem[Glazebrook et al.<2004>]{Glazebrook04}
Glazebrook, K. et al. 2004, Nature, 430, 181

\bibitem[Guzman et al.<1997>]{Guzman97}
Guzman, R, et al. 1997, ApJ, 489, 559

\bibitem[Harker et al.<2006>]{Harker06}
Harker G., Cole S., Helly J., Frenk C. S., Jenkins A., 2006, MNRAS, 367, 
1039 
\bibitem[H{\"a}ring \& Rix<2004>]{Haring04} 
H{\"a}ring, N., \& Rix, H.-W. 2004, ApJ, 604, L89 

\bibitem[Helly et al.<2006>]{Helly06}
Helly J. C., Cole S., Frenk C. S., Baugh C. M., Benson A. J., Lacey C., 2003, 
MNRAS, 338, 903 

\bibitem[Hopkins et al.<2007>]{Hopkins07} 
Hopkins, P.~F., Bundy, K., Hernquist, L., \& Ellis, R.~S.  2007, ApJ, 659, 976 

\bibitem[Kauffman et al.<1999>]{Kauffmann99}
Kauffmann~G., Colberg~J.~M., Diaferio,~A. \& White,~S.~D.~M. 1999, MNRAS, 303, 188

\bibitem[Kauffman et al.<2003>]{Kauffmann03}
Kauffmann~G. et al., 2003, MNRAS, 341, 54

\bibitem[Kennicutt<1983>]{Kennicutt83}
Kennicutt, R.~C., Jr. 1983, ApJ, 272, 54 

\bibitem[Kennicutt<1998>]{Kennicutt98}
Kennicutt~R.~C., 1998, Annual Review of Astronomy \& Astrophysics, 36, 189

\bibitem[Kere\v{s} et al.<2005>]{Keres}
Kere\v{s}~D., Katz~N., Weinberg~D.~H., Dav\'e~R., 2005, MNRAS, 363, 2

\bibitem[Kitzbichler \& White<2006>]{Kitzbichler06}
Kitzbichler~M.~G. \& White~S.~D.~M., 2006, MNRAS, 366, 858

\bibitem[Kitzbichler \& White<2007>]{KW07} 
Kitzbichler~M.~G., \& White,~S.~D.~M. 2007, MNRAS, 376, 2 

\bibitem[Lawrence et al.<2007>]{Lawrence07} 
Lawrence~A., et al. 2007, MNRAS, 379, 1599 

\bibitem[Le F{\`e}vre et al.<2005>]{leFevre05}
Le F{\`e}vre, O. et al., 2005, 439, 845

\bibitem[Martin \& Kennicutt<2001>]{Martin01} 
Martin, C.~L., \& Kennicutt, R.~C., Jr. 2001,ApJ, 555, 301 

\bibitem[McCarthy et al.<2008>]{McCarthy08}
McCarthy~I.~G., Frenk~C.~S., Font~A.~S., Lacey~C.~G., Bower~R.~G., Mitchell~N.~L., Balogh~M.~L., Theuns~T., 2008, MNRAS, 383, 593

\bibitem[McKee \& Ostriker<1977>]{McKee77}
McKee~C.~F. \& Ostriker~J.~P., 1977, ApJ,  218, 148

\bibitem[Navarro, Frenk, \& White<1995>]{Navarro95}
Navarro~J.~F., Frenk~C.~S. \& White~S.~D.~M.,1995, MNRAS, 275, 720

\bibitem[Pozzetti et al.<2007>]{Pozzetti07}
Pozzetti~L, et al. 2007, AA, 474, 443

\bibitem[Scoville et al.<2007>]{Scoville07} 
Scoville, N., et al. 2007, ApJS, 172, 38 

\bibitem[Somerville \& Primack <1999>]{Somerville99}
Somerville~R.~S. \& Primack~J.~R. 1999, MNRAS, 310, 1087

\bibitem[Somerville et al.<2004>]{Somerville04}
Somerville~R.~S., Lee~K., Ferguson~H.~C., Gardner~J.~P., Moustakas~L~A., \& Giavalisco~M. 2004, ApJ, 600, L171

\bibitem[Springel, White \& Jenkins<2005>]{Springel05a}
Springel, V., White, S. D. M. \& Jenkins, A., 2005, Nat, 435, 629

\bibitem[Springel et al. <2005>]{Springel05b} Springel, V., Di 
Matteo, T., \& Hernquist, L. 2005, MNRAS, 361, 776 

\bibitem[Sutherland \& Dopita<1993>]{Sutherland93}
Sutherland~R.~S. \& Dopita~M.~A., 1993, ApJ,  88, 253

\bibitem[van Dokkum et al.<2006>]{vanDokkum06}
van Dokkum, P. G. et al. 2006, ApJ, 638, L59

\bibitem[Willmer et al.<2006>]{Willmer06}
Willmer, C. N. A et al. 2006, ApJ, 647, 853

\bibitem[Wolf et al.<2003>]{Wolf03}
Wolf~C., Meisenheimer~K., Rix~H.-W., Borch~A.,
 Dye~S. \& Kleinheinrich~M., 2003, AA, 401, 73
 
\end{thebibliography}
\end{document}